\documentclass[amsmath,amssymb,aps,prl,superscriptaddress,twocolumn]{revtex4-2}
\pdfoutput=1
\usepackage{graphicx}
\usepackage{dcolumn}
\usepackage{bm}
\usepackage[usenames]{xcolor}
\usepackage{hyperref} 
\usepackage{subfigure}
\usepackage{amsmath}
\usepackage{amssymb}
\usepackage{relsize}
\usepackage[group-separator={,}]{siunitx}

\DeclareMathOperator{\relu}{ReLU}

\begin{document}

\title{Correlator Convolutional Neural Networks: An Interpretable Architecture for
Image-like Quantum Matter Data}
\author{Cole Miles}
\affiliation{Department of Physics, Cornell University, Ithaca, NY 14853, USA}
\author{Annabelle Bohrdt}
\affiliation{Department of Physics, Harvard University, Cambridge, MA 02138, USA}
\affiliation{Department of Physics and Institute for Advanced Study, Technical University of Munich, 85748 Garching, Germany}
\affiliation{Munich Center for Quantum Science and Technology (MCQST), 80799 M\"unchen, Germany}
\author{Ruihan Wu}
\affiliation{Department of Computer Science, Cornell University, Ithaca, NY 14853, USA}
\author{Christie Chiu}
\affiliation{Department of Physics, Harvard University, Cambridge, MA 02138, USA}
\affiliation{Department of Electrical Engineering, Princeton University, Princeton, NJ 08540, USA}
\affiliation{Princeton Center for Complex Materials, Princeton University, Princeton, NJ 08540, USA}
\author{Muqing Xu}
\affiliation{Department of Physics, Harvard University, Cambridge, MA 02138, USA}
\author{Geoffrey Ji}
\affiliation{Department of Physics, Harvard University, Cambridge, MA 02138, USA}
\author{Markus Greiner}
\affiliation{Department of Physics, Harvard University, Cambridge, MA 02138, USA}
\author{Kilian Q.\ Weinberger}
\affiliation{Department of Computer Science, Cornell University, Ithaca, NY 14853, USA}
\author{Eugene Demler}
\affiliation{Department of Physics, Harvard University, Cambridge, MA 02138, USA}
\author{Eun-Ah Kim}
\affiliation{Department of Physics, Cornell University, Ithaca, NY 14853, USA}

\begin{abstract}
Machine learning models are a powerful theoretical tool for analyzing data from quantum simulators, in which results of experiments are sets of snapshots of many-body states. Recently, they have been successfully applied to distinguish between snapshots that can not be identified using traditional one and two point correlation functions. Thus far, the complexity of these models has inhibited new physical insights from this approach. Here, using a novel set of nonlinearities we develop a network architecture that discovers features in the data which are directly interpretable in terms of physical observables. In particular, our network can be understood as uncovering high-order correlators which significantly differ between the data studied. We demonstrate this new architecture on sets of simulated snapshots produced by two candidate theories approximating the doped Fermi-Hubbard model, which is realized in state-of-the art quantum gas microscopy experiments. From the trained networks, we uncover that the key distinguishing features are fourth-order spin-charge correlators, providing a means to compare experimental data to theoretical predictions. Our approach lends itself well to the construction of simple, end-to-end interpretable architectures and is applicable to arbitrary lattice data, thus paving the way for new physical insights from machine learning studies of experimental as well as numerical data.
\end{abstract}


\maketitle

Increasingly, image-like experimental data from quantum systems promises to offer greater insight into the physics of correlated quantum matter \cite{bakr_quantum_2009, mazurenko_cold-atom_2017, zhang_machine_2019, stinson_imaging_2018, iwasawa_development_2017, moler_imaging_2017}. However, the traditional framework of condensed matter physics \cite{ashcroft_solid_1976} lacks principled approaches for analyzing such image-like data and connecting the data with theoretical insights. Without such techniques, the connection between theory founded on simple fundamental principles and image-like data is at best a one-way street where theory can produce approximate images that only partially resemble the real data. Hence often the validity of a theory has been advocated for based on the degree of resemblance in select features of theory-based and real data.

There have been growing efforts to adopt data science tools that have proved effective at recognizing every-day objects for objective analysis of image-like data on quantum matter \cite{zhang_machine_2019, rem_identifying_2019, bohrdt_classifying_2019}. The key idea is to use the ability of neural networks to express and model functions to learn key features found in the image-like data in an objective manner. However, there are two central challenges to this approach. First, the ``black box'' nature of neural networks is particularly problematic when it comes to scientific applications, where it is critical that the outcome of the analysis is based on scientifically correct reasoning \cite{zhang_machine_2018}. The second challenge unique to scientific application of supervised machine learning (ML) approaches is the shortage of real training data. Hence the community has generally relied on simulated data for training \cite{zhang_machine_2019, bohrdt_classifying_2019,ghosh_one-component_2020}. However, it has not been clear whether the neural networks trained on simulated data properly generalize to experimental data. The path to surmounting both of these issues is to obtain some form of interpretability in our models. To date, most efforts at interpretable machine learning on scientific data have relied on manual inspection and translation of learned features from training standard architectures \cite{wetzel_machine_2017,casert_interpretable_2019, bluecher_towards_2020}. Instead, here we propose an entirely new approach designed from the ground-up to automatically learn information that is meaningful within the framework of physics.

The need for a principled data-centric approach is particularly great and urgent in the case of synthetic matter experiments such as quantum gas microscopy (QGM) \cite{bakr_quantum_2009}, ion traps \cite{jurcevic_direct_2017}, and Rydberg atom arrays \cite{labuhn_tunable_2016, bernien_probing_2017}. While our technique is generally applicable, in this work we focus on QGM, which enables researchers to directly sample from the many-body density matrix of strongly correlated quantum states that are simulated using ultra-cold atoms. With the quantum simulation of the fermionic Hubbard model finally reaching magnetism \cite{mazurenko_cold-atom_2017} and the strange metal regime \cite{brown_bad_2019, koepsell_microscopic_2020}, QGM is poised to capture a wealth of information on this famous model that bears many open questions and is closely linked to quantum materials. However, the real-space snapshots QGM measures are a fundamentally new form of data resulting from a direct projective measurement of a many-body density matrix as opposed to a thermal expectation value of observables. While this means richer information is present in a full dataset, little is known about how to efficiently extract all the information. When it comes to the questions regarding the enigmatic underdoped region of the fermionic Hubbard model, the challenge is magnified by the fact that fundamentally different theories can predict QGM data with seemingly subtle differences within standard approaches \cite{chiu_string_2019, koepsell_microscopic_2020}.

\begin{figure*}[ht!]
	\includegraphics[width=\linewidth]{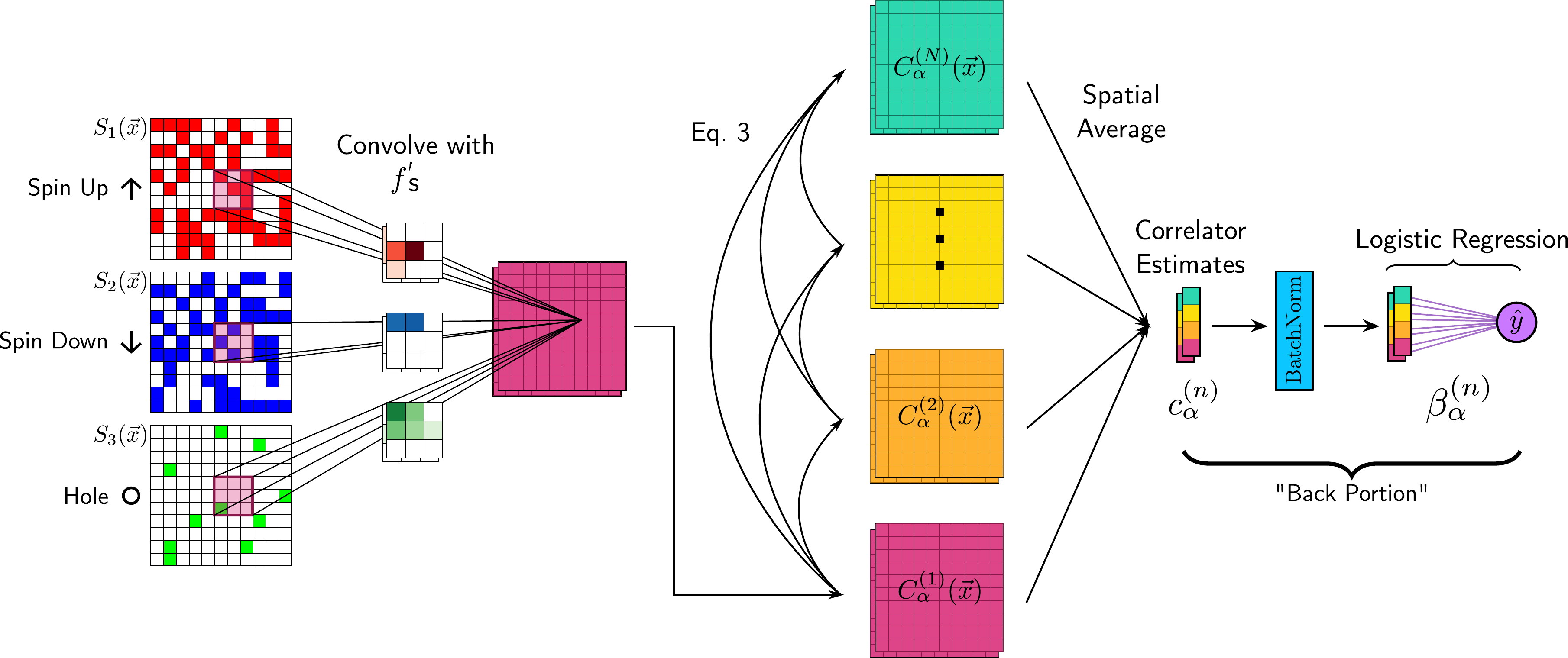}
	\caption{The construction of our Correlation Convolutional Neural Network, shown here with two learnable filters $(M=2)$. The input is a three-channel image: $S_{1}(\vec{x}) = n_{\uparrow}(\vec{x}), S_{2}(\vec{x}) = n_{\downarrow}(\vec{x}), S_{3}(\vec{x}) = n_{\text{hole}}(\vec{x})$. The image is first convolved with learned filters $f_{\alpha}$ to produce a set of convolutional maps $C^{(1)}_\alpha(\vec{x})$. Maps containing information about higher-order local correlations can then be recursively constructed using the lower-order maps, truncating at some order $N$. Spatially averaging these maps produces features $c_\alpha^{(n)}$ which in expectation are equal to weighted sums of correlators found within the corresponding convolutional filter. These features are normalized to zero mean and unit variance by a BatchNorm layer, then used by a logistic classifier with coefficients $\beta_\alpha^{(n)}$ to produce the final output $\hat{y}$.}
	\label{fig:arch}
\end{figure*}
\begin{figure}[h!]
	\subfigure[\label{fig_string_cartoon}]{%
		\includegraphics[width=0.45\columnwidth]{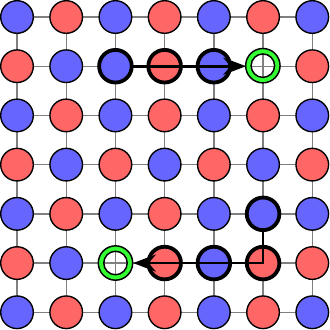}
	}\hfill
	\subfigure[\label{fig_piflux_cartoon}]{%
		\includegraphics[width=0.45\columnwidth]{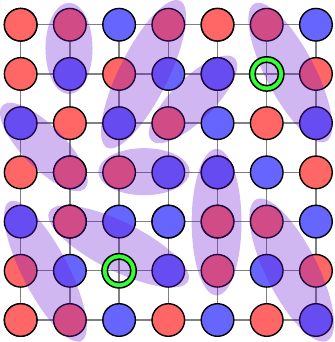}
	}
	\caption{A cartoon depicting the features of two candidate theories approximating the low-$T$, low-doping limit of the Fermi-Hubbard model. (a) Geometric string theory, showing two geometric strings in the presence of an antiferromagnetic background. Note that the propagation of the doped holes creates parallel line segments of aligned spins, perpendicular to the direction of the hole propagation. (b) $\pi$-flux theory, which describes a spin liquid of singlet pairs.}
	\label{fig:cartoon}
\end{figure}

In this letter, we develop Correlator Convolutional Neural Networks (CCNNs), a novel architecture with a set of nonlinearities designed to produce features that are directly interpretable in terms of correlation functions in image-like data (see Figure~\ref{fig:arch}). Following training of this architecture, we employ \textit{regularization path analysis} \cite{tibshirani_least_2004} to rigorously identify the features that are critical in the CCNN's performance. We apply this powerful combination of CCNNs and regularization path analysis to simulated and experimental QGM data of the under-doped Fermi Hubbard model. Following this, we discuss the new insights we gain regarding the hidden signatures of two theories, geometric string theory \cite{grusdt_parton_2018} and $\pi$-flux theory \cite{marston_large-_1989, wen_theory_1996}, as well as application to non-spin-resolved experimental data.

The Hubbard model of fermionic particles on a lattice is a famous model that bears many open questions and is closely linked to quantum materials such as high-temperature superconductors. The model Hamiltonian is given by 
\begin{equation}
	\mathcal{H} = -t\sum_{\sigma=\uparrow,\downarrow}\sum_{\langle i, j\rangle} (\hat{c}^\dagger_{i, \sigma} \hat{c}_{j, \sigma} + \text{h.c.}) + U \sum_{i} \hat{n}_{i, \uparrow} \hat{n}_{i, \downarrow}
\end{equation}
where the first term describes the kinetic energy associated to electrons hopping between lattice sites, and the second term describes an on-site repulsion between electrons. 
At half-filling, and in the limit $U \gg t$, the repulsive Hubbard model maps to the Heisenberg antiferromagnet \cite{auerbach_interacting_1994}. However, the behavior of the model as the system is doped away from half-filling is not as well-understood.
Several candidate theories exist which attempt to describe this regime, including geometric string theory \cite{grusdt_parton_2018} and $\pi$-flux theory \cite{marston_large-_1989, wen_theory_1996}. These theories are conceptually very distinct, but at low dopings measurements in the occupation basis do not differ enough in simple conventional observables such as staggered magnetization or two-point correlation functions to fully explain previous ML success \cite{bohrdt_classifying_2019} in discrimination (see SM Sec.~S.IV). Nevertheless, there are more subtle hidden structures involving more than two sites \cite{chiu_string_2019} which are noticeable. In the ``frozen spin approximation'' \cite{grusdt_microscopic_2019}, geometric string theory predicts that the motion of the holes simply displaces spins backwards along the path the hole takes. Hence the propagation of the doped hole will tend to produce a ``wake'' of parallel line segments of aligned spins in its trail (see Fig.~\ref{fig_string_cartoon}).
Meanwhile, the $\pi$-flux theory describes a spin liquid of singlet pairs, where it is more difficult to conceive of characteristic structures (see Fig.~\ref{fig_piflux_cartoon}).

Current QGM experiments are able to directly simulate the Fermi-Hubbard model, obtaining one or two-dimensional occupation snapshots sampled from the thermal density matrix $\rho \sim e^{-\beta \mathcal{H}}$ prescribed by the model \cite{mazurenko_cold-atom_2017}. However, currently our experiment can only resolve a single spin species at a time, leaving all other sites appearing as empty. This is not a fundamental limitation of QGM experiments and in principle, complete spin and charge readout is possible \cite{salomon_direct_2019, koepsell_imaging_2019}. As we aim to learn true spin correlations, in this work we use primarily simulated snapshots at doping $\delta=0.09$ sampled from the geometric string and $\pi$-flux theories using Monte Carlo sampling techniques under periodic boundary conditions (see SM Sec.~S.I).

We point out that in the context of this paper, when referring to two models as different, we do not imply that they are fundamentally distinct, in the sense that they can not be connected smoothly without encountering a singularity in the partition function. Rather, this is a practical question: we have two or more mathematical procedures for generating many-body snapshots based on variational wavefunctions, Monte-Carlo sampling, or any other theoretical approach. Our goal is to develop a ML algorithm that separates snapshots based on which procedure they are more likely to come from and, most importantly, the algorithm should provide information about which correlation functions are most important for making these assignments.

To learn how to distinguish these two theories we propose a novel neural network architecture, Correlation Convolutional Neural Networks (CCNNs), schematically shown in Fig.~\ref{fig:arch}.
The input to the network is an image-like map with 3-channels $\{S_k(\vec{x}) | k\!\!=\!\!1,2,3\}$, where $S_1(\vec{x}) = n_{\uparrow}(\vec{x}), S_2(\vec{x}) = n_{\downarrow}(\vec{x}), S_3(\vec{x}) = n_{\text{hole}}(\vec{x})$. Since the models we consider are restricted to the singly-occupied Hilbert space, this input only takes on values $0$ or $1$. From this input, the CCNN constructs nonlinear ``correlation maps'' containing information of local spin-hole correlations up to some order $N$ across the snapshot. This operation is parameterized by a set of learnable 3-channel filters, $\{f_{\alpha, k} | \alpha\!\!=\!\!1, \cdots, M\}$ where $M$ is the number of filters in the model.
The maps for the given filter $\alpha$ are defined as:
\begin{align}
    C^{(1)}_\alpha(\vec{x}) &= \sum_{\vec{a}, k} f_{\alpha, k}(\vec{a}) S_k(\vec{x} + \vec{a}) \nonumber \\
    C^{(2)}_\alpha(\vec{x}) &= \sum_{(\vec{a}, k) \ne (\vec{b}, k')} f_{\alpha, k}(\vec{a}) f_{\alpha, k'}(\vec{b}) S_k(\vec{x} + \vec{a}) S_{k'}(\vec{x} + \vec{b}) \nonumber \\
                     &\vdots \label{eq_Fn} \\
    C^{(N)}_\alpha(\vec{x}) &= \sum_{(\vec{a}_1, k_1) \ne \dots \ne (\vec{a}_N, k_N)} \prod_{j=1}^N f_{\alpha, k_j}(\vec{a}_j) S_{k_j}(\vec{x} + \vec{a}_j). \nonumber
\end{align}
Here $\vec{a}$ runs over the convolutional window of the filter $\alpha$. Traditional convolutional neural networks employ only the first of these operations, alternating with some nonlinear activation function such as $\tanh$ or $\text{ReLU}(x) = \max(0, x)$. The issue with these typical nonlinear functions is that they mix all orders of correlations into the output features, making it difficult to disentangle what exactly traditional networks measure. In contrast, each order of our nonlinear convolutions $C^{(n)}_\alpha(\vec{x})$ are specifically designed to learn $n$-site semi-local correlations in the vicinity of the site $\vec{x}$, which appear as patterns in the convolutional filters $f_{\alpha}$. Note that the sums in Eq.~\ref{eq_Fn} exclude any self-correlations to aid interpretability. During training, a CCNN tunes the filters $f_{\alpha, k}(\vec{a})$ such that correlators characteristic of the labeled theory are amplified while others are suppressed. To aid interpretation, we force all filters to be positive $f_{\alpha, k}(\vec{a}) \ge 0$ by taking the absolute value before use on each forward pass.

A direct computation of the nonlinear convolutions following Eq.~\ref{eq_Fn} up to order $N$ requires $\mathcal{O}((KP)^N)$ operations per site, where $P$ is the number of pixels in the window of the filter and $K$ is the number of species of particles. However, we can use the following recursive formula which we prove in the Supplement, Section S.II\ref{supp_sec_proof}:
\begin{align}
    C^{(n)}_\alpha(\vec{x}) &= \nonumber \\
    \frac{1}{n} \sum_{l=1}^n & (-1)^{l-1} \left(\sum_{\vec{a}, k} f_{\alpha, k}(\vec{a})^l S_k(\vec{x} + \vec{a})^l \right) C^{(n-l)}_\alpha(\vec{x}), \label{eq_recur}
\end{align}
where all powers are done pixelwise \footnote{Note that for our input, $S^l = S$.}, and we define $C_\alpha^{(0)}(\vec{x}) = 1$. This improves the computational complexity to $\mathcal{O}(N^2KP)$ while also allowing us to leverage existing highly-optimized GPU convolution implementations. Use of this formula leads to a ``cascading'' structure to our model similar to \cite{roheda_conquering_2020}, as seen in Fig.\ \ref{fig:arch}. First, the input $S$ is convolved with filters $f_\alpha$ to produce the first-order maps $C_\alpha^{(1)}$. Using Eq.~\ref{eq_recur}, these first order maps can be used to construct second order maps $C_\alpha^{(2)}$, and onwards until the model is truncated at some order $N$. Since the Hamiltonians being studied are translation-invariant, we then obtain estimates of correlators from these correlation maps by simple spatial averages to produce $c_{\alpha}^{(n)} = \frac{1}{N_{\text{sites}}} \sum_{\vec{x}} C^{(n)}_{\alpha}(\vec{x})$. Concatenating these correlator estimates results in an $NM$-dimensional feature vector $\vec{c} = \{c_\alpha^{(n)}\}$. 

In the back portion of a CCNN (see Fig.~\ref{fig:arch}), the feature vector $\vec{c}$ is normalized using a BatchNorm layer \cite{ioffe_batch_2015}, then used by a logistic classifier which produces the classification output
$	\hat{y}(\vec{c};\vec{\beta},\epsilon)= [1 + \exp(-\vec{\beta}\cdot \vec{c} + \epsilon)]^{-1}$
where $\vec{\beta} = \{\beta_\alpha^{(n)}\}$ and $\epsilon$ are trainable parameters. If $\hat{y} < 0.5$, the snapshot is classified as $\pi$-flux, and otherwise it is classified as geometric string theory. The $\beta_\alpha^{(n)}$ coefficients are central to the interpretation of the final architecture, as they directly couple the correlator features $c_\alpha^{(n)}$ to the output.
For training, we use L1 loss in addition to the standard cross-entropy loss, \textit{i.e.} 
\begin{equation}
	L_{\text{train}}(y, \hat{y}) \equiv -y \log \hat{y} -(1-y)\log(1 - \hat{y}) + \gamma \sum_{\alpha, k, \vec{a}} f_{\alpha, k}(\vec{a}),
\end{equation}
where $y = \{0, 1\}$ is the label of the snapshot, and $\gamma$ is the L1 regularization strength. The role of the L1 loss is to bias the filter patterns to be simple by turning off pixels which are unnecessary (see SM Sec.~S.I)\footnote{During the completion of this work, we learned that a similar approach (L1 regularization on the convolutional filters) was taken in Ref.\cite{bluecher_towards_2020}}.

We fix the number of filters $M$ and the maximum order of the non-linear convolutions $N$, a hyper-parameter specific to CCNN, by systematically observing the training performance. We found that two filters gives sufficient performance while allowing for simple interpretation. Hence we consider two filters, i.e., $M=2$ in the rest of the paper. For the maximum order of non-linear convolution $N$ we found the performance to rapidly increase with increase in $N$ up to $N=4$, past which performance plateaus. (see SM Sec.~S.I.)
Hence we fix $N=4$ in the rest of the paper. Additionally, we limit our investigation to $3\times 3$ convolutional filters. With the architecture of the CCNN so-fixed we found the performance of this minimalistic model to be comparable with a more complex traditional CNN architecture \cite{bohrdt_classifying_2019} (see SM Sec.~S.I). 

\begin{figure}[h!]
	\includegraphics[width=\columnwidth]{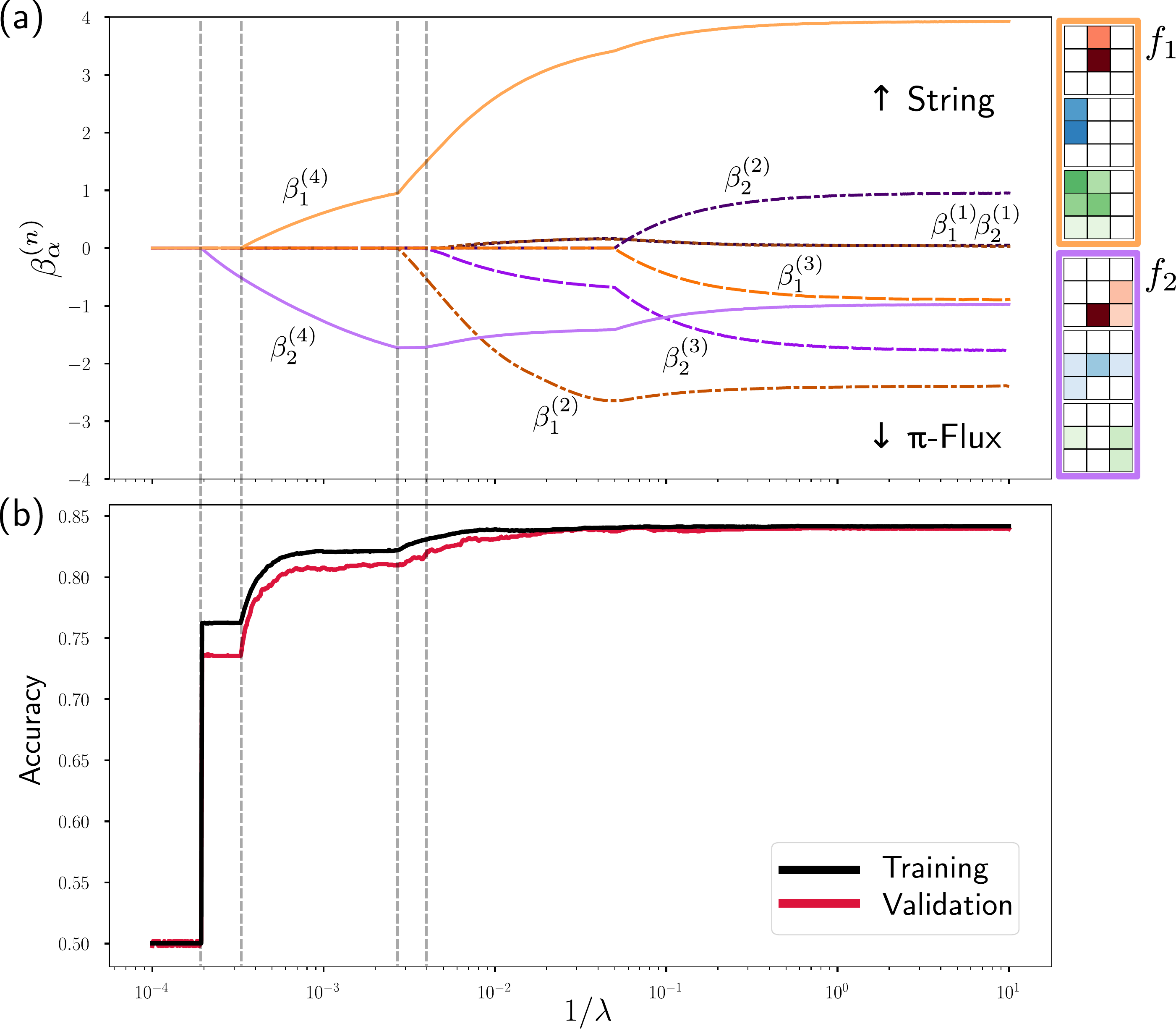}
	\caption{(a) The regularization path of $\beta_\alpha^{(n)}$ coefficient values traced out by two learned filters as a function of the inverse regularization strength $1/\lambda$.
	Positive and negative signs of $\beta_\alpha^{(n)}$ are associated with geometric string and $\pi$-flux labels respectively.
	(b) The accuracies of the model at each point of the regularization path in (a) on both the training dataset, as well as a held-back validation dataset which is unseen by the model during training.}
	\label{fig_reg_path}
\end{figure}

After a CCNN is trained, we fix the convolutional filters $f_\alpha$ and move on to a second phase to interpret what it has learned. We first determine which features are the most relevant to the model's performance
by constructing and analyzing regularization paths \cite{tibshirani_least_2004} to examine the role of the logistic coefficients $\beta_\alpha^{(n)}$. We apply an L1 regularization loss to these $\beta_\alpha^{(n)}$ and re-train the back portion of the model (see Fig.~\ref{fig:arch}) using a new loss function:
\begin{equation}
	L_{\text{path}}(y, \hat{y}) \equiv -y \log \hat{y} -(1-y)\log(1 - \hat{y}) + \lambda \sum_{\alpha, n} |\beta_\alpha^{(n)}|,
\end{equation}
where $\lambda$ is the regularization strength. Now the regularization by $\lambda$ penalizes the use of coefficients $\beta_\alpha^{(n)}$ and the corresponding use of the features $c_\alpha^{(n)}$.
This results in an optimization trade-off between minimizing the classification loss and attempting to keep $\beta_\alpha^{(n)}$ at zero, where the relative importance of these terms is tuned by $\lambda$. At large $\lambda$, the loss is minimized by keeping all $\beta_\alpha^{(n)}$ at zero, resulting in a 50\% classification accuracy due to the model always predicting a single class. As $\lambda$ is slowly ramped down, eventually the ``most important'' coefficient $\beta_\alpha^{(n)}$ will begin to activate, due to the decrease in classification loss surpassing the increase in the activation loss. As these coefficient couple the correlator features $c_\alpha^{(n)}$ to the prediction output, this process offers clear insight into which features are the most relevant.

We show a typical regularization path analysis in Fig.~\ref{fig_reg_path}, where the filters $f_\alpha$ of a trained model are shown in the inset. The activation of each coefficient $\beta_\alpha^{(n)}$ is tracked while tuning down the regularization strength $\lambda$ (increasing $1/\lambda$). The resulting trajectories in Fig.~\ref{fig_reg_path}(a) show that the 4th order correlator features, $c_1^{(4)}$ and $c_2^{(4)}$ are most significant for the CCNN's decision making since $\beta_1^{(4)}$ and $\beta_2^{(4)}$ are the two first coefficients to activate. Furthermore, parallel tracking of the accuracy in Fig.~\ref{fig_reg_path}(b) shows that the activation of these features results in large jumps in the classification accuracy, comprising almost all of the network's predictive power. While the details of the paths vary between training runs, we find robust dominance of fourth-order correlations as the first features to be activated to give the majority of the network's performance.

\begin{figure}
    \includegraphics[width=0.98\columnwidth]{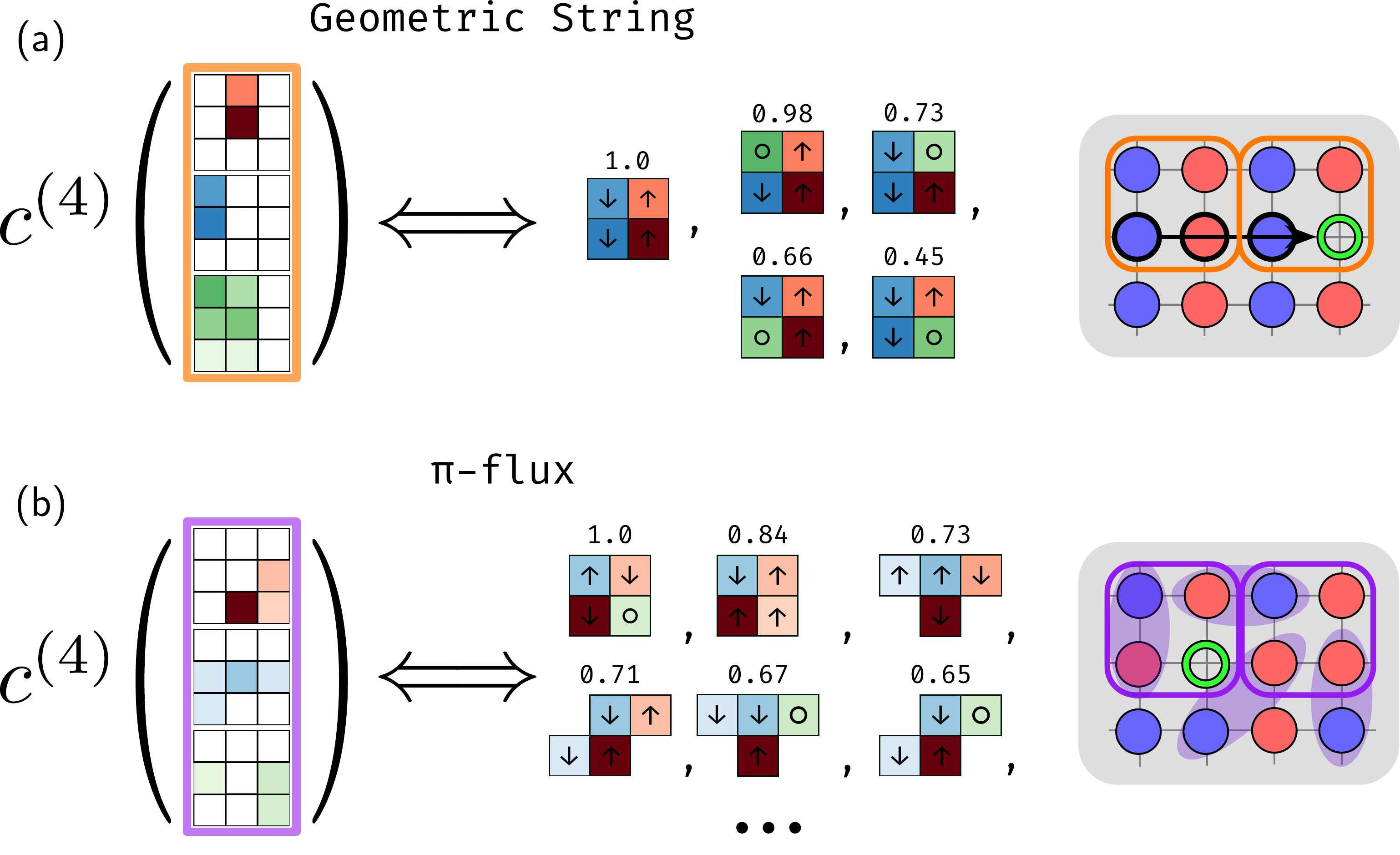}
	\caption{The highest-weight terms of Eq.~\ref{eq_Fn} when constructing correlator features $c^{(4)}_1, c^{(4)}_2$ from the discovered convolutional filter patterns $f_1, f_2$. Each feature $c^{(4)}_\alpha$ measures a weighted sum of the correlators drawn on the right-hand side. Weights shown here are normalized such that the largest correlator from each filter has weight $1.0$.}
	\label{fig_expansion}
\end{figure}

Now that we know the fourth-order correlations are the important features, we look at which physical correlators are being measured by the features $c^{(4)}_\alpha$ by simply inspecting 4-pixel patterns made from high-intensity pixels from each channel of the learned filters, as we show in Fig.~\ref{fig_expansion}. Comparing these patterns with the depiction of the two candidate theories, we can understand why these correlators measured by the two filters are indeed prominent motifs. Specifically, the $2\times 2$ correlators in the fourth-order feature of the filter associated to the geometric string theory (Fig.~\ref{fig_expansion}(a)) are easily recognizable in the ``wake'' and the termination of a string. These discovered correlations are in agreement with those examined in Ref.~\cite{koepsell_imaging_2019}, which found pronounced spin anti-correlations induced on the spins located on the diagonal adjacent to a mobile chargon. Meanwhile, the $2 \times 2$ motifs in the filter learned to represent the $\pi$-flux theory (Fig.~\ref{fig_expansion}(b)) are either a single spin-flip or a simple placement of a hole into an AFM background.
It is evident that this CCNN is learning the fingerprint correlations of geometric string theory, recognizing the $\pi$-flux theory instead from fluctuations which are uncharacteristic of the string picture. Furthermore, a subset of learned patterns that are not obvious from the simple cartoons can be used as additional markers to detect the states born out of the two theoretical hypotheses in experiment (see SM Sec.~S.IV for more detail).

It is important to note that the above insights relied on the fact that our CCNN's structure can be understood as measured collections of correlators. Although the regularization path analysis can be applied to any architecture, the typical non-linear structures of off-the-shelf CNNs inhibit direct connections between the dominant filters and physically meaningful information~\cite{khatami_visualizing_2020}. In SM Sec~S.VI we present how interpretation of the architecture of Ref.~\cite{bohrdt_classifying_2019} can be attempted following similar steps as above. Since the fully connected layer contains tens of thousands of parameters, after training we show that we can reduce this layer to a simple spatial averaging to attempt interpretation, with no loss in performance (see SM Sec.~S.V). The reduced architecture with a single ``feature'' per convolutional filter, similar to the architecture of Ref.~\cite{khatami_visualizing_2020}, is trained, after which we fix the filters for the regularization path analysis. We can clearly determine which filters produce the important features, but it is unclear what these features are actually measuring due to the $\relu$ nonlinearity. However, without any nonlinearity the architecture only achieves close to 50\% performance. This failure to enforce simplicity on traditional architectures shows the importance of designing an architecture which measures physically meaningful information from the outset.

The ML method presented in this paper considers short-range multi-point correlations functions (up to 3 lattice sites in both $x$ and $y$ directions), but does not include long-range two-point correlations needed for identifying spontaneous symmetry breaking. Two considerations motivate this choice: i) Current experiments with the Fermi-Hubbard model are done in the regime where  correlations involving charge degrees of freedom are not expected to exceed a few lattice constants due to thermal fluctuations. ii) The energy of systems with local interactions, such as the Fermi-Hubbard model, is primarily determined by short-range correlations. We note, however, that the current method can be extended to include longer range correlations either by expanding the size of the filters used in Eq.~\ref{eq_Fn}, or by using dilated convolutions.

To summarize, we proposed a new neural network architecture that is inherently interpretable as measuring sets of multi-site correlators. We then applied this architecture to the supervised learning problem of distinguishing two theoretical hypotheses for the doped Hubbard model: $\pi$-flux theory and geometric string theory. Employing a regularization path analysis technique on these trained CCNN architectures, 
we found that four-site correlators deriving from the learned filters hold the key fingerprints of geometric string theory. A subset of these four-site motifs fit into what is expected from the wake of a propagating hole in an antiferromagnetic background. The remaining four-site motifs which go beyond our existing intuition offer new insight into the problem. 

The broad implications of CCNN-based machine learning for analysis and acquisition of image-like data are twofold.
Firstly, the revelation of specific high-order correlations as defining features of target states can guide experimental design. Specifically for QGM, our CCNN approach found fourth-order correlators carry the key signature distinguishing geometric string theory and $\pi$-flux snapshots. Our discovered patterns are consistent with a recent report on the importance of a specific fifth-order correlator found with a targeted manual analysis of simulated data~\cite{bohrdt_dominant_2020}. These observations indicate that future experiments should target measurement of higher-order correlations, as also recommended in \cite{bohrdt_classifying_2019, koepsell_microscopic_2020}. Secondly, CCNN can reveal any critical gaps between simulated and experimental data for neural-network based hypothesis testing. In particular, we found that for the data we had access to, experimental uncertainties on the actual doping and temperature due to the lack of spin-resolution allowed the CCNN to focus solely on the doping level rather than meaningful correlation functions (see SM Sec.~S.VI). For successful hypothesis testing, spin-resolved QGM datasets, which are just becoming accessible \cite{salomon_direct_2019, koepsell_imaging_2019}, will be necessary. With growing access to spatially resolved image-like data in quantum matter, we anticipate our CCNN approach can bring the power of neural networks to the design and analysis of image-like data that organically fits into the tradition of physics.

\noindent
{\bf Acknowledgements.} 
We thank Fabian Grusdt and Andrew Gordon Wilson for insightful discussions during the completion of this work. CM acknowledges that this material is based upon work supported by the U.S.\ Department of Energy, Office of Science, Office of Advanced Scientific Computing Research, Department of Energy Computational Science Graduate Fellowship under Award Number DE-SC0020347. AB, RW, KW, ED, E-AK acknowledge support by the National Science Foundation through grant No.~OAC-1934714. AB acknowledges funding by Germany's Excellence Strategy - EXC-2111 - 390814868.

{\bf Disclaimer.}
This report was prepared as an account of work sponsored by an agency of the United States Government. Neither the United States Government nor any agency thereof, nor any of their employees, makes any warranty, express or implied, or assumes any legal liability or responsibility for the accuracy, completeness, or usefulness of any information, apparatus, product, or process disclosed, or represents that its use would not infringe privately owned rights. Reference herein to any specific commercial product, process, or service by trade name, trademark, manufacturer, or otherwise does not necessarily constitute or imply its endorsement, recommendation, or favoring by the United States Government or any agency thereof. The views and opinions of authors expressed herein do not necessarily state or reflect those of the United States Government or any agency thereof.

\bibliographystyle{apsrev4-2}
\bibliography{qgm-interpretable}

\onecolumngrid

\newpage
\appendix

\renewcommand\thefigure{S.\arabic{figure}}
\setcounter{figure}{0}
\renewcommand\theequation{S\arabic{equation}}
\setcounter{equation}{0}
\setcounter{table}{0}
\renewcommand{\thetable}{S\arabic{table}}


\section{Supplementary Materials}

\subsection{S.I Training Details} \label{sec_supp_training}

\paragraph{Training data generation}

Equilibrium snapshots of both models -- $\pi$-flux theory and geometric string theory -- can be sampled straightforwardly, as discussed in \cite{bohrdt_classifying_2019} and \cite{chiu_string_2019}. We here briefly summarize the corresponding sampling techniques.

Geometric string theory only makes a statement about how the doped model deviates from half-filling. The spin-background at half-filling here is given by sampling snapshots of the Heisenberg model at finite temperature using quantum Monte Carlo techniques. We sample snapshots for a $40\times 40$ system with periodic boundary conditions and then cut out a $16 \times 16$ observation region from each snapshot. 
For a given doping level, we then insert the corresponding number of holes into each snapshot at random positions, where holes cannot sit on the same site. The geometric string theory provides a distribution of string lengths for a given temperature \cite{grusdt_microscopic_2019,chiu_string_2019}. For each hole, we thus sample a string length from the string length distribution and move the hole by hand for a corresponding number of sites through the spin background in random directions while displacing the spins along its path. 

Snapshots from $\pi$-flux theory are generated using standard Metropolis Monte Carlo sampling of the Gutzwiller projected thermal density matrix of the mean-field Hamiltonian \cite{baskaran_resonating_1987}
\begin{equation}
\begin{split}
\hat{\mathcal{H}}_{\rm MF} = &-\frac{1}{2} J^* \sum_{\vec{i} \in A} \sum_\sigma \left( e^{i\theta_0} \hat{c}_{\vec{i},\sigma}^\dagger \hat{c}_{\vec{i}+\vec{x},\sigma} + e^{-i\theta_0} \hat{c}_{\vec{i},\sigma}^\dagger \hat{c}_{\vec{i}+\vec{y},\sigma} + h.c. \right) 
\\
&-\frac{1}{2} J^* \sum_{\vec{i} \in B} \sum_\sigma \left( e^{-i\theta_0} \hat{c}_{\vec{i},\sigma}^\dagger \hat{c}_{\vec{i}+\vec{x},\sigma} + e^{i\theta_0} \hat{c}_{\vec{i},\sigma}^\dagger \hat{c}_{\vec{i}+\vec{y},\sigma} + h.c. \right).
\label{eqHMF}
\end{split}
\end{equation}
Here, $\vec{i} \in A(B)$ denotes lattice sites $\vec{i}$ which are part of the A(B) sublattice and $\hat{c}_{\vec{i},\sigma}^{(\dagger)}$ is the annihilation (creation) operator of a fermion with spin $\sigma$. The mean-field Hamiltonian describes a system with staggered flux $\pm \Phi = \pm 4 \theta_0$ and effective hopping amplitude $J^*$. In particular, we consider $\pi$-flux states with $\theta_0=\pi/4$.
We simultaneously sample the occupation in momentum and real space. 
The real and momentum space configurations are denoted as $|\tilde{\alpha}_{\vec{r}}\rangle$ and $| \alpha_{\vec{k}} \rangle$, respectively. In momentum space, the two spin species are treated separately, such that two fermions of opposite spin can occupy the same momentum state. 
In real space, two fermions with opposite spin cannot occupy the same site, thus directly implementing the Gutzwiller projection. In any given real space configuration $|\tilde{\alpha}_{\vec{r}}\rangle$, each site is therefore either empty or occupied with a spin up or a spin down fermion. 
The mean field Hamiltonian \eqref{eqHMF} can be readily diagonalized in momentum space.
For each momentum space configuration $| \alpha_{\vec{k}} \rangle$, we thus directly obtain an energy $E(\alpha_{\vec{k}})$ and from that the corresponding thermal weight. We use the Metropolis Monte Carlo algorithm \cite{Gros1989} to sample Gutzwiller projected real space snapshots $| \tilde{\alpha}_{\vec{r}} \rangle$ according to the probability distribution
\begin{equation}
p_{\beta}(\tilde{\alpha}_{\vec{r}},\alpha_{\vec{k}}) = Z^{-1} e^{-\beta E(\alpha_{\vec{k}})} | \langle \tilde{\alpha}_{\vec{r}} | \alpha_{\vec{k}} \rangle |^2.
\end{equation}
The overall energy scale $J^*$ of this model is treated as a free parameter which is fit so that the nearest-neighbor spin correlators match with those of geometric string theory at half-filling. We sample snapshots of size $16\times 16$ with periodic boundary conditions.

Our dataset consists of $10000$ sampled snapshots from each theory, $9000$ of which go into the training set which is seen by the network, with the other $1000$ being reserved for validation. This makes our training set size $18000$, and our validation set size $2000$. The exact partitioning of snapshots into these sets is determined by a random seed we call the ``split seed''. 

\paragraph{Training Procedure}
As described in the main text, training is done in two phases. The first is done in PyTorch \cite{paszke_pytorch_2019}, in which the full model, including both the convolutional filters and the logistic classifier, is trained using the \texttt{ADAM} optimization algorithm. The resulting model from this process can be used as-is for classification. However, for an interrogation of which features are most important, regularization paths such as those in Fig.~\ref{fig_reg_path} are produced in a second phase. Here, the convolutional filters are held fixed and only the back logistic classifier is trained multiple times over a wide range of L1 $\lambda$ coefficients. This phase is done in Scikit-Learn \cite{pedregosa_scikit-learn_2011} due to its extremely efficient logistic regression routines.

\begin{table}[h]
	\centering
	\begin{tabular}{|l|c|}
		\hline
		Optimizer & \texttt{Adam} \\
		Adam $\beta_1$ & 0.9 \\
		Adam $\beta_2$ & 0.999 \\
		Adam $\epsilon$ & $1\times 10^{-8}$ \\
		Learning Rate & 0.005 \\
		LR Schedule & \texttt{CosineAnnealingLR} \\
		Batch Size & 1024 \\
		L1 Coefficient, $\gamma$ & 0.005 \\
		Epochs & 1000 \\
		Number of Filters & 2 \\
		Split Seed & 1111 \\
		Batching Seed & 4444 \\
		\hline
	\end{tabular}
	\caption{Hyperparameters used to train the model used to produce the results of Figs.\ \ref{fig_reg_path} and \ref{fig_expansion}. Random seeds are included to allow full reproducibility using the code which will be made public.}
\end{table}

During the first phase, the optimization process attempts to minimize the loss:

\begin{equation}
	L_{\text{train}}(y, \hat{y}) \equiv -\sum_{i} y_i \log \hat{y}_i + \gamma \sum_{\alpha} ||f_\alpha||_1 
\end{equation}
where the first term is the standard cross-entropy loss used for classification tasks, and the second term is an L1 (or LASSO) regularization which has the effect of driving the unimportant components of the convolutional weights $f_\alpha$ to zero.

To allow for simple interpretation of the resulting filters, we strictly limit the $f_\alpha$ to take on only positive values. This can either be done by replacing the $f_\alpha$ in place with their absolute value after each gradient update, or by simply taking the absolute value every time a forward pass is done. This was found to incur a $\sim2\%$ accuracy loss, which we find acceptable in order for easier interpretation. In contrast, forcing the filter weights to be positive seems to entirely halt the learning process for traditional CNN architectures.

The loss during the second phase is similar:

\begin{equation}
	L_{\text{path}}(y, \hat{y}) \equiv -\sum_{i} y_i \log \hat{y}_i + \lambda ||\vec{\beta}||_1
\end{equation}
with the notable difference that the LASSO regularization is now being applied to the logistic weights rather than the convolutional filters.

As shown in Fig.\ \ref{fig_drawing_corr}, we utilize a BatchNorm \cite{ioffe_batch_2015} layer intermediate between the nonlinear convolutions and the logistic classifier, without the additional learnable affine transformation typically used as we find these introduce additional complexity without much benefit for our problem. During training, this layer simply normalizes the features produced from each minibatch to be zero mean and unit variance to allow for easier classification. During validation, the layer uses exponential running estimates of the mean and variance for normalization rather than the minibatch statistics. We empirically found this layere to be essential to creating a well-performing architecture, with the hypothesis that this is related to the different scales that each nonlinear feature tends to exist at. Normalization brings all of the features to the same relative scale, allowing the classifier to have an easier time detecting distributional shifts.

However, there does exist an unintended interaction between L1 regularization and BatchNorm. Due to the normalization process, the architecture is invariant to an overall scaling of the filter weights. Meanwhile, our intended goal of using L1 was to drive ``unimportant'' pixels to zero. In a sense, the network can do this for ``free'' since it can scale the filter weights without any loss in performance. In practice however, we find that the L1 loss still does bias the network towards having lower complexity filters. However the relationship between the $\lambda$ parameter and the number of pixels activated is not always simple - sometimes increasing $\lambda$ will result in more pixels activated. While the interaction between L2 regularization and BatchNorm is well understood \cite{van_laarhoven_l2_2017}, the authors of this work are unaware of any similar understanding with L1 regularization. A solution to this issue is still a desired feature.

\paragraph{Symmetrization}

\begin{figure}
	\centering
	\includegraphics[width=0.75\linewidth]{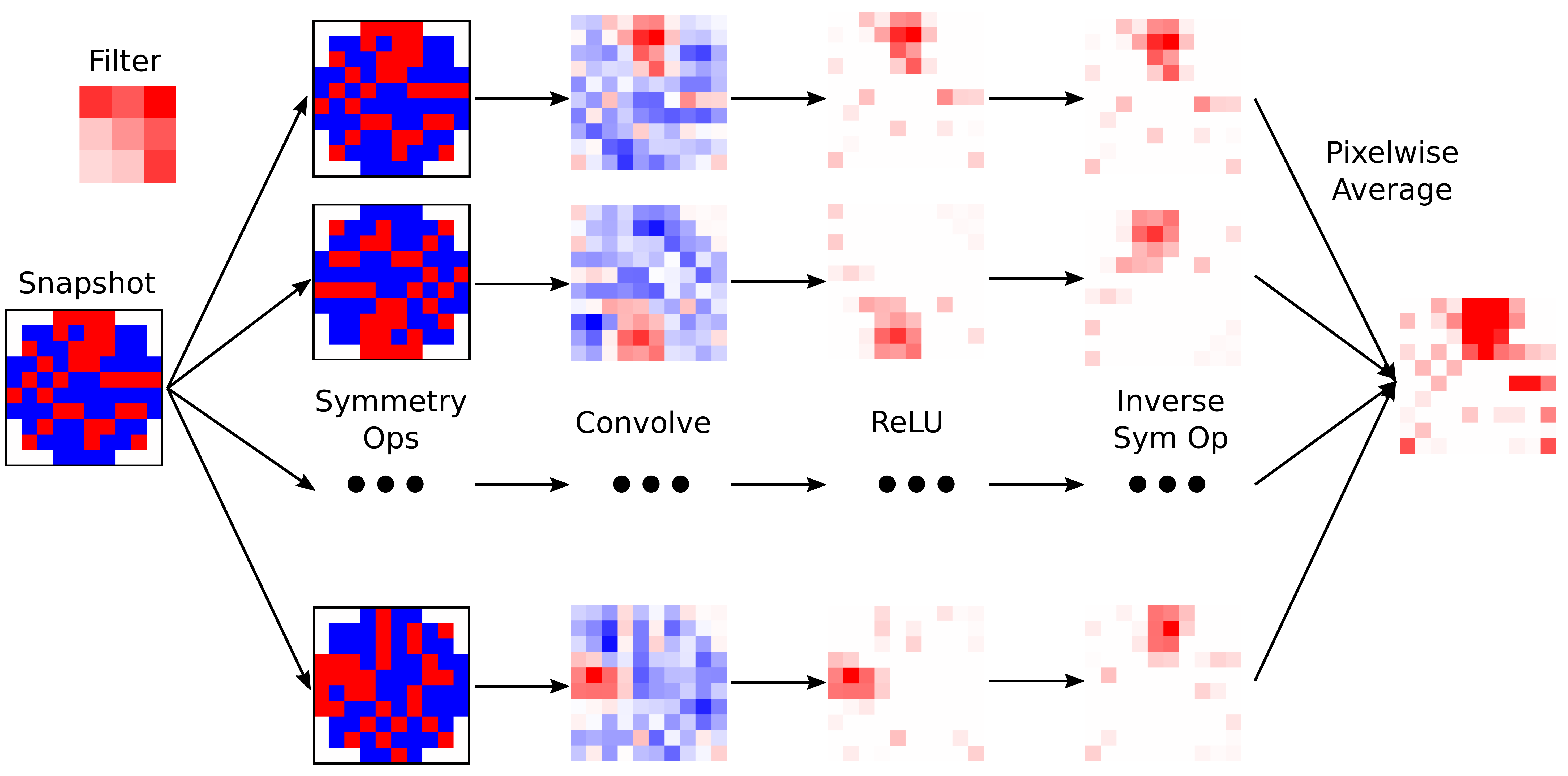}
	\caption{Diagram demonstrating $D^8$ symmetric convolutions, as done in a traditional CNN architecture. Note that this diagram does not reflect the exact construction of our architecture, but only serves to show the symmetrization procedure.}
	\label{fig_supp_symconv}
\end{figure}

One factor leading to the overparameterization of standard CNNs is that to reach peak accuracy, they need to explicitly learn multiple symmetry-equivalent versions of spin patterns. To achieve the same effect without requiring the duplication of filters, we use a $D^8$ symmetry-equivariant form of the convolutional operation as introduced in \cite{dieleman_exploiting_2016}. A visual explanation of the operation as performed in a standard CNN pipeline can be seen in Fig.\ \ref{fig_supp_symconv}.

Modification to suit our architecture is simple, following the steps described in \cite{dieleman_exploiting_2016} to extend this idea to arbitrary models. Before any operation is applied, a ``symmetric slicing'' operation is done which stacks extra rotated/flipped copies of the input into the batch dimension. The rest of the operations in the architecture are applied as usual to the entire batch. Then, before feeding the final features into the logistic classifier, a ``symmetric pooling'' operation applies the correct inverse symmetry operations to each copy of the input, then averages across them. This entire block of operations then forms features which are equivariant to the desired symmetries of the input. If these features are then spatially averaged, they instead form invariants (in which case the aforementioned inverse symmetry operations are not strictly needed). For fair comparison, every model examined in this work had this symmetrization applied.

\paragraph{Performance Measurements}

\begin{figure}[h!]
	\subfigure[Running-max performance of our architecture compared at different orders we construct the model. We compare against a traditional CNN architecture using $\relu$ as the nonlinear function.\label{fig_supp_performance_order}]{%
		\includegraphics[width=0.45\columnwidth]{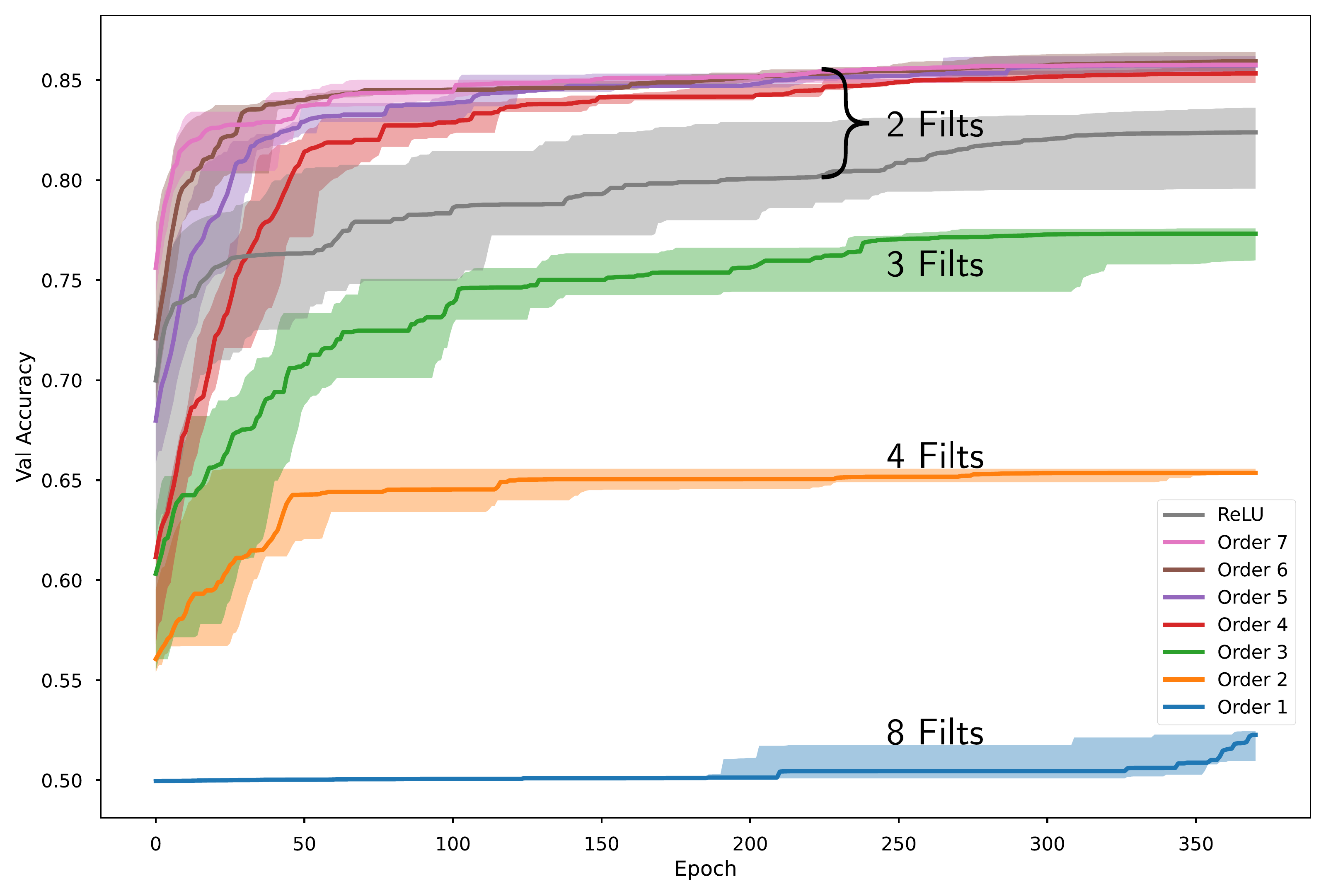}
	}\hfill
	\subfigure[Performance of a fourth-order architecture containing different numbers of convolutional filters.\label{fig_supp_performance_nfilts}]{%
		\includegraphics[width=0.49\columnwidth]{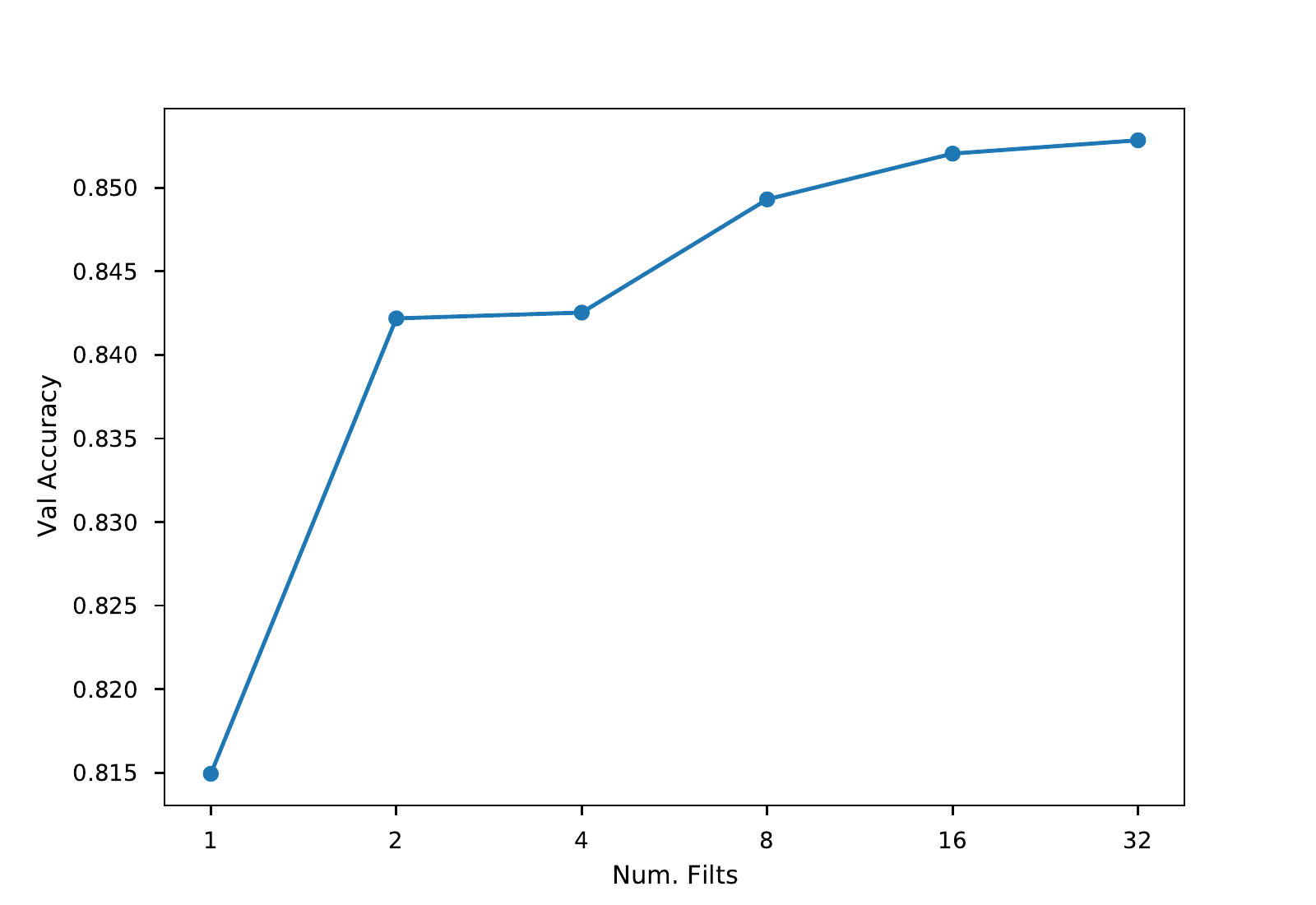}
	}
	\caption{Performance results. The solid lines indicate the median of five independently-trained models, while the shaded regions show the min-max spread across these models.}
	\label{fig_supp_performance}
\end{figure}

In Fig.\ \ref{fig_supp_performance_order}, we show the performance of our architecture at various orders to which the model is constructed, compared against a traditional CNN architecture using $\relu$ as the nonlinearity. (For details, see the ``reduced architecture'' described in Sec~S.V). We also have compared to the architecture of \cite{bohrdt_classifying_2019}, adapted to accept three-channel snapshots as input, though it is difficult to control the overfitting even with strong regularization. Meanwhile, our architecture does not show signs of significant overfitting even in absence of regularization due to its small parameterization. Out of all of our trials, no tested CNN has outperformed our CCNN models on the validation dataset.

Each curve shown is labeled with the number of filters that model contains; we increase the number of filters as the order of the model decreases to keep the total number of features relatively constant, for a fair comparison. The solid lines show the running-max (over all previous epochs) of the median validation accuracy achieved between five independent training runs on the same train-val split of the data, but with different parameter initializations and batching order, while the shaded regions shown the min-max spread across these models. Note that, to avoid unfairly biasing the higher-order networks, the models shown here are not trained with the L1 regularization on the filter weights.

\begin{figure}[h!]
	\centering
	\subfigure[Filters learned from a variety of two-filter models \label{fig_supp_twofilts}]{%
    	\includegraphics[width=0.75\linewidth]{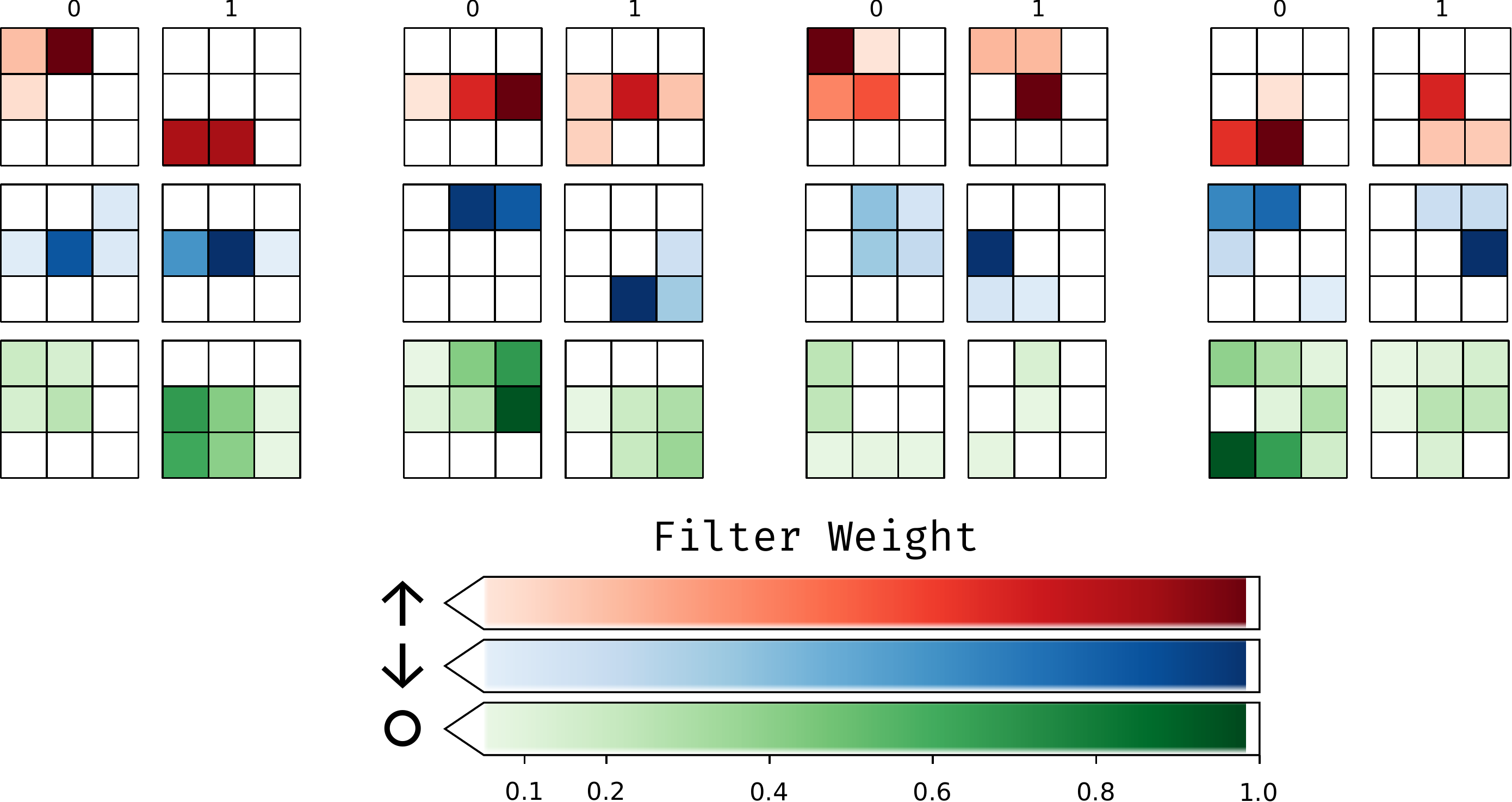}
	}\\
	\subfigure[Filters learned from a variety of single-filter models \label{fig_supp_singlefilt}]{%
	    \includegraphics[width=0.4\linewidth]{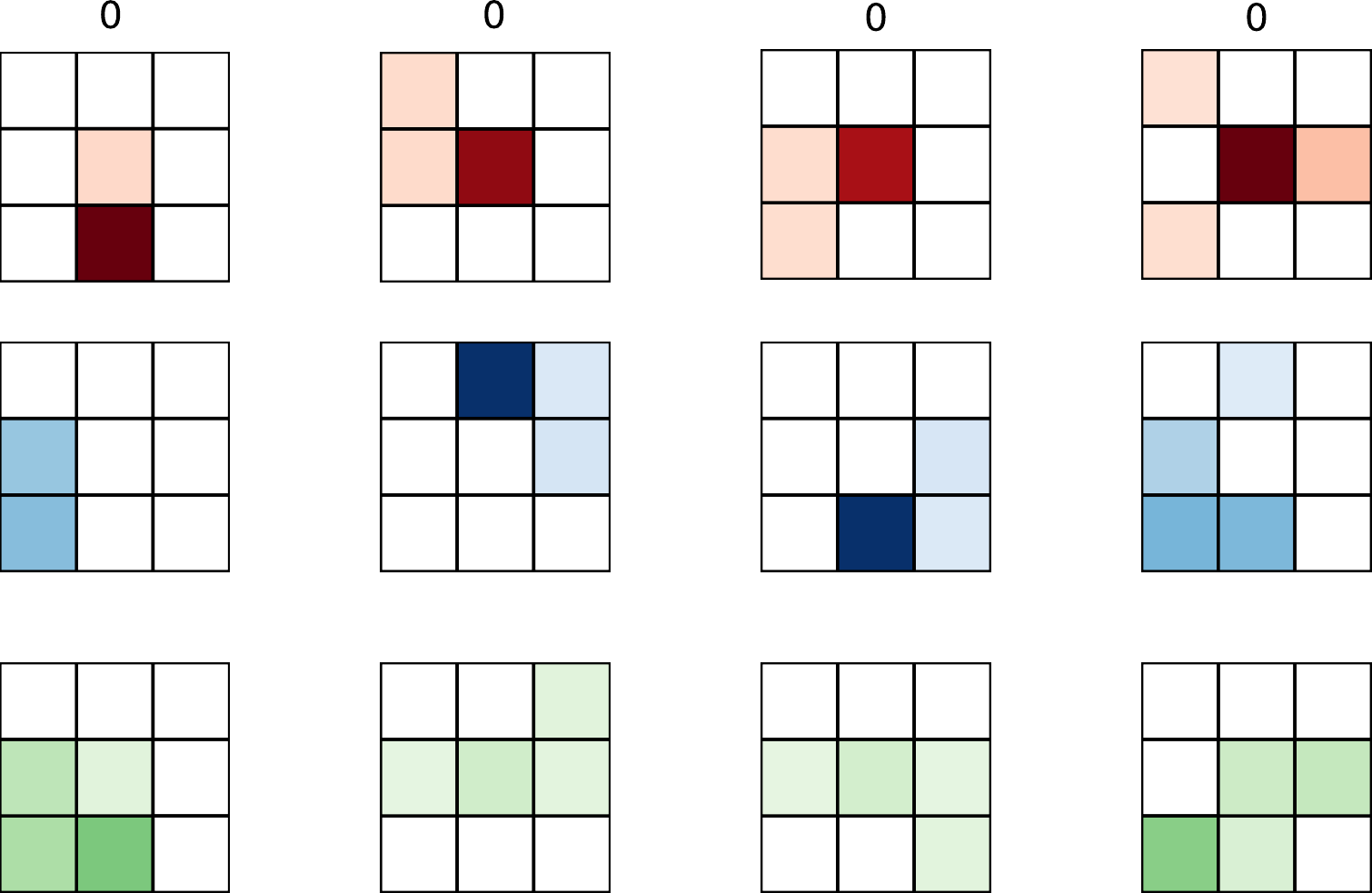}
	}
	\caption{Examples of filters learned from independently trained models.}
	\label{fig_supp_examples}
\end{figure}

We see that the performance of the architecture rapidly increases as a function of the order to which the model is constructed, plateauing past fourth order. At this order, we consistently match performance with traditional CNN architectures, even with only two convolutional filters and dramatically fewer learnable parameters. We find performance plateauing past fourth order to be a general behavior, independent of the number of filters, regularization strength, etc. This indicates that fifth-order and higher correlations provide no new statistically significant information between the two datasets, at least at the snapshot sizes we use.

In Fig.\ \ref{fig_supp_performance_nfilts}, we show the final trained performance of our architecture as a function of the number of filters used. For these measurements, L1 regularization is turned off, as it may prevent additional filters from being used at all. Interestingly, we find that we can get close to optimal performance with just a single convolutional filter, with performance quickly plateauing past this.

We find that the parallel spin and L-shaped patterns shown in the main text are generally robust features learned by the architecture. The ``interlocking-L'' pattern is extremely robust, with some variant occurring in nearly every trained model. While the parallel spin pattern does not appear in every trained model, it does seem to be the second most common pattern. In Fig.\ \ref{fig_supp_examples}, we show examples of models trained on the same data with different random seeds controlling the initialization. We note that while the exact filter pattern varies between training runs, the dominant subpatterns tend to match between all of runs.

\clearpage
\newpage

\subsection{S.II Proof of Eq.\ \ref{eq_recur}} \label{supp_sec_proof}

For simplicity, in this section we will use subscripts to represent spatial indexing, and forgo a channel index. I.e. $S_{i}$ represents the value of the input at site $i$. Since information from different filters is never mixed, it is sufficient to consider this operation with a single filter $f_{a}$ where $a$ indexes the sites within the convolutional window. Furthermore, define $x_a \equiv f_a S_{i+a}$. Our goal is then to show, with the definition

\begin{equation}
	C^{(n)} \equiv \sum_{a_1 \ne \dots \ne a_n} \prod_{j=1}^n x_{a_j},
\end{equation}
that the following recursive formula holds:
\begin{equation}
	C^{(n)} = \frac{1}{n} \sum_{l=1}^n (-1)^{l-1} \left(\sum_{a} x_{a}^l \right) C^{(n-l)}, \label{eq_supp_recur}
\end{equation}
with the definition that $C^{(0)} = 1$.

This can be confirmed by direct substitution and unwrapping the sum term-by-term. The $l = 1$ term of the sum reads

\begin{equation}
	(l=1) \hspace{1cm} \left(\sum_{a} x_a\right) \sum_{a_1 \ne \dots \ne a_{n-1}} \prod_{j=1}^{n-1} x_{a_j}
\end{equation}

If we imagine expanding this as a sum of products of $n$ variables, there will exist two types of terms: those where all $x$'s in the term are unique, and those where $x_a$ from the first sum equals exactly one of the $x_{a_j}$. Each of the former is overcounted by a factor of $n$, and each of the latter by a factor of $n-1$:

\begin{align}
	(l=1) \hspace{1cm} \left(\sum_{a} x_a\right) \sum_{a_1 \ne \dots \ne a_{n-1}} \prod_{j=1}^{n-1} x_{a_j} &= n\sum_{a_1 \ne \dots \ne a_n} \prod_{j=1}^n x_{a_j} + (n-1) \sum_{a_1 \ne \dots \ne a_{n-1}} x_{a_1}^2 \prod_{j=2}^{n-1} x_{a_j} \nonumber \\
	&= nC^{(n)} + (n-1)\sum_{a_1 \ne \dots \ne a_{n-1}} x_{a_1}^2 \prod_{j=2}^{n-1} x_{a_j} \label{eq_supp_term1}
\end{align}

The rest of the terms in the $l$ sum of Eq.\ \ref{eq_supp_recur} serve solely to cancel out the extraneous terms on the right. We can see that the $l=2$ term splits into two pieces, similar as to how the $l=1$ term did:

\begin{align}
	(l=2) \hspace{1cm} -\left(\sum_{a} x_a^2\right) \sum_{a_1 \ne \dots \ne a_{n-2}} \prod_{j=1}^{n-2} x_{a_j}
	&= -(n-1)\sum_{a_1 \ne \dots \ne a_{n-2}} x_{a_1}^2 \prod_{j=2}^{n-2} x_{a_j} - (n-2)\sum_{a_1 \ne \dots \ne a_{n-3}} x_{a_1}^3 \prod_{j=2}^{n-3} x_{a_j} \label{eq_supp_term2}
\end{align}

So, adding together the $l=1$ and $l=2$ terms cancels the second sum on the right hand side of Eq.\ \ref{eq_supp_term1} which contains terms involving $x_a^2$, but introduces another extraneous sum of terms involving $x_a^3$. In general, the $l$\textsuperscript{th} term expands to

\begin{equation}
	(-1)^{l-1}\left( (n-l+1) \sum_{a_1 \ne \dots \ne a_{n-l}} x_{a_1}^{l} \prod_{j=2}^{n-l} x_{a_j}  + (n-l) \sum_{a_1 \ne \dots \ne a_{n-l-1}} x_{a_1}^{l+1} \prod_{j=2}^{n-l-1} x_{a_j}\right).
\end{equation}

From this expansion, we can see that the left piece of the $l$\textsuperscript{th} summand cancels the right piece of the $l-1$\textsuperscript{th} summand. This expansion continues to unzip up until the $l=n$ term, $(-1)^{n-1}\sum_a x_a^n$, in which the right piece is zero. Hence, once the sum is fully unzipped, the only term remaining is the left piece of the $l=1$ summand, which is exactly $C^{(n)}$.

This result still holds for multi-channel images and filters. To restore a channel index, all of the above equations will remain true with the transformations
\begin{equation}
    x_{a} \equiv f_{a} S_{i+a} \rightarrow x_{k, a} \equiv f_{k, a} S_{k, i+a} \hspace{1cm} \sum_{a_1 \ne \cdots \ne a_n} \rightarrow \sum_{(a_1, k_1) \ne \cdots \ne (a_n, k_n)}
\end{equation}
where $k = \{1, 2, \dots K\}$ runs over the number of channels $K$ in the input snapshot.

Using Eq.\ \ref{eq_supp_recur}, we can compute each of these $C^{(n)}$ in order, efficiently utilizing the results of previous computations to only require $\mathcal{O}(N^2 KP)$ operations per site total. The coefficients in parentheses can be seen to be the result of taking the convolution of the $l$\textsuperscript{th} power of the convolutional filter with the $l$\textsuperscript{th} power of the occupancy snapshot, taken pixelwise. We can save on an extra bit of computation (though not changing the overall complexity) if the system is fermionic, in which case $S^{l} = S$ for arbitrary $l \ge 1$.

\subsection{S.III Understanding the Behavior of Convolutional Activation Maps under Nonlinearities}

Here we will take a moment to understand how nonlinearities applied to convolutional activation maps can be understood as extracting local correlations similar to the learned filter pattern. Consider an input snapshot $S_x$, containing some observable at each location $x$ (we can also imagine $x$ as a multi-index including a channel dimension, to allow for multiple observables at each location). A convolution with some filter $f_a$ indexed within a window by $a$ then produces the map $C_x = f_a S_{x+a}$, where the sum over $a$ is implied.

Within this section, we will define the expectation of variables as:
\begin{equation}
    \langle \cdot \rangle \equiv \mathbb{E}_{S \sim e^{-\beta \mathcal{H}}}[\cdot],
\end{equation}
that is, the average value of the variable when $S$ is sampled from the thermal density matrix of some Hamiltionian $\mathcal{H}$. Before applying any nonlinearity, the features we measure on average are just:

\begin{equation}
    \langle C_x \rangle = f_a \langle S_{x+a} \rangle.
\end{equation}
Hence, what we would measure is simply a weighted sum of the average values of the observables -- in our case, the occupations. This would be true for any linear map applied to the data. If we instead apply an analytic nonlinear function $\sigma$ to the activation map, Taylor expand about $f = 0$, and then take the expectation we instead measure:

\begin{equation}
    \langle\sigma(C_x)\rangle = \sigma(0) + \sigma'(0) f_a \langle S_{x+a}\rangle + \sigma''(0) f_a f_b \langle S_{x+a} S_{x+b}\rangle + \sigma'''(0) f_a f_b f_c \langle S_{x+a} S_{x+b} S_{x+c}\rangle + \dots \label{eq_supp_nonlin_expansion}
\end{equation}

We can see that each order of the expansion is incorporating information about the same-order correlation functions involving sites contained within the convolutional window. Hence, we can say that CNNs using traditional nonlinearities measure spatial correlations in the data at \textit{all} orders which fit into the convolutional window.

\begin{figure}[h!]
    \centering
    \includegraphics[width=\columnwidth]{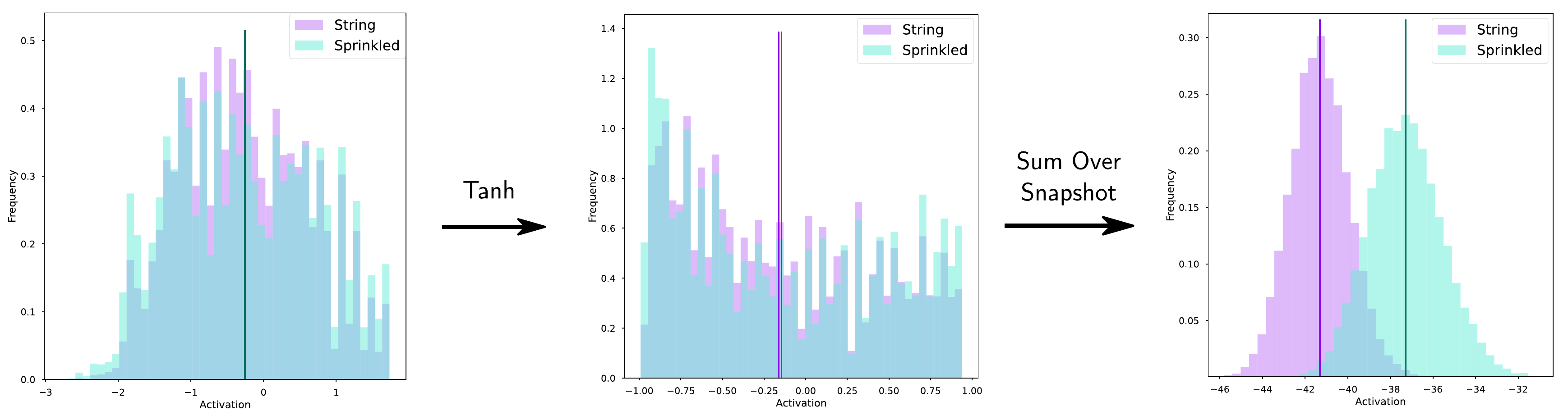}
    \caption{Histograms tracking the distribution of activations in the network after various operations.}
    \label{fig_supp_activation_hists}
\end{figure}

We demonstrate this on a simple example in Fig.\ \ref{fig_supp_activation_hists}. Here, we have trained a simple CNN identical to that in Fig.\ \ref{fig_supp_reduced_arch}, but using the $\tanh$ nonlinearity rather than $\relu$, and only with a single filter. The model is trained to distinguish between snapshots sampled from the geometric string theory, and from the ``sprinkled holes'' theory described in \cite{chiu_string_2019}, where holes are simply randomly placed into snapshots sampled from the same antiferromagnetic Heisenberg model as used for string theory snapshot generation. The first histogram in the figure shows the distribution of $C_x$ when the trained model is applied to snapshots from each theory. As both theories have identical occupation statistics the means of each distribution are essentially identical, implying that $C_x$ directly is not a useful feature for classification. However, we can notice that the \textit{variances} $\sim \langle C_x^2\rangle $ related to two-point correlations, are slightly different.

The next histogram shows the distribution of $\tanh(C_x)$, where we can now see that the means of each distribution have separated very slightly. An alternate way to understand Eq.~\ref{eq_supp_nonlin_expansion} is the statement that the mean of $\sigma(C_x)$ is related to all higher moments of the $C_x$ distribution. The separation in means of $\tanh(C_x)$ is small on the level of individual pixels, but we can see that after a summation, the final distributions of $\sum_x \tanh(C_x)$ results are well separated allowing for simple classification. While we may not be able to perform an expansion for nonanalytic functions such as $\relu$, it is this property of mixing moments of the $C_x$ distribution that allows nonlinearities to introduce higher-order correlations into the learned features.

The central idea behind CCNNs derives from these observations: we can directly use powers of the convolutional activation map to extract out correlations of different orders. Our nonlinear operations in Eq.~\ref{eq_Fn} are simple polynomials of the inputs, designed such that self-correlations are removed for simpler direct interpretation.

\subsection{S.IV Exact Measurements} \label{sec_supp_measurements}

To confirm that our ML models are indeed finding true physical features, we have explicitly calculated correlator estimates for both of the training datasets. 

\begin{figure}[h!]
    \centering
    \subfigure[Staggered magnetization]{
        \includegraphics[width=0.4\linewidth]{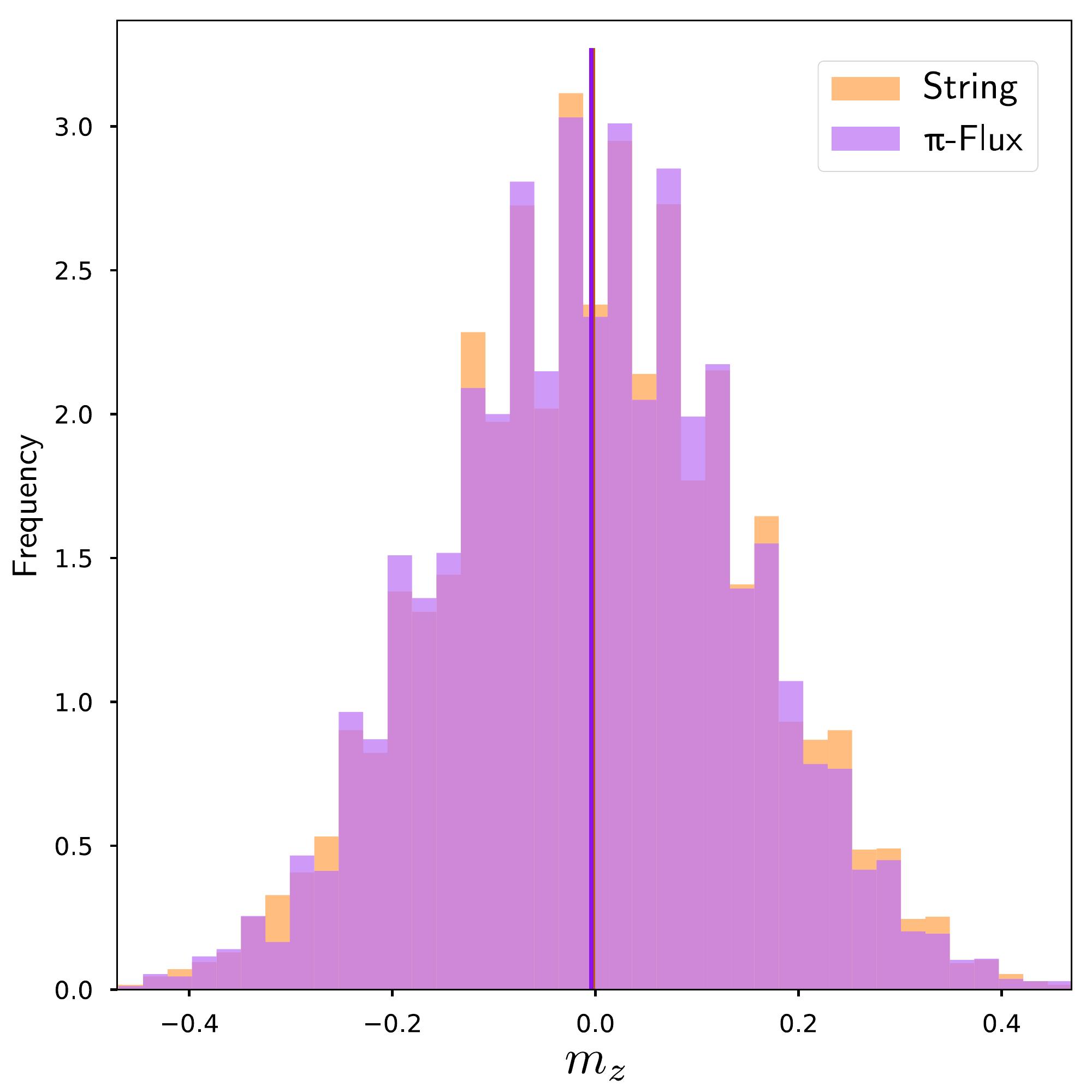}
    }\hspace{1cm}
    \subfigure[Nearest-neighbor sign-corrected correlator]{
        \includegraphics[width=0.4\linewidth]{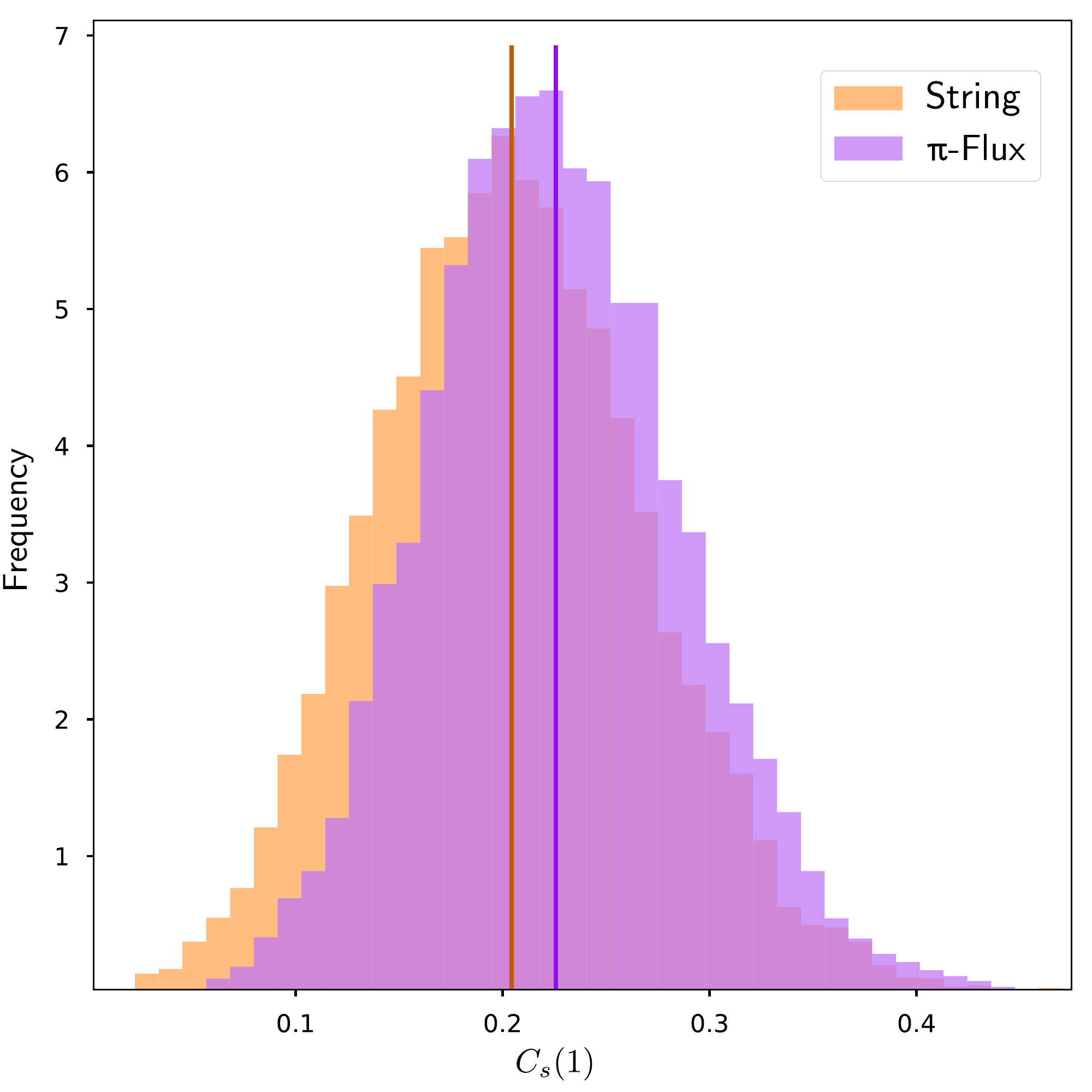}
        \label{fig_supp_simple_nncorr}
    }
    \caption{Histograms of ``simple'' observables measured from single snapshots of each theory, with each histogram scaled to integrate to one. Vertical lines denote the mean of each distribution.}
    \label{fig_supp_simple_meas}
\end{figure}

We first measure some examples of ``simple'' observables at the $\delta = 0.09$ doping level studied in this work. We define $s_{ij} = +1, -1, 0$ if a spin up, spin down, or hole, respectively, lives at site $(i, j)$. We measure the staggered magnetization,
\begin{equation}
    m_z = \sum_{i, j} (-1)^{i+j} s_{i,j}
\end{equation}
and the sign-corrected nearest neighbor spin-spin correlator,
\begin{equation}
    C_s(1) = -\left(\frac{\langle s_{i,j} s_{i+1, j}\rangle + \langle s_{i,j} s_{i, j+1} \rangle - \langle s_{i, j} \rangle \langle s_{i+1, j}\rangle - \langle s_{i, j} \rangle\langle s_{i, j+1} \rangle}{2}\right)
\end{equation},
with results shown in Fig.~\ref{fig_supp_simple_meas}.

We see that the staggered magnetization is nearly indistinguishable, though the nearest-neighbor spin correlator does show some minor deviation between the theories. However, this discrepancy is hardly enough to explain the $>80\%$ classification accuracy of ours and previous \cite{bohrdt_classifying_2019} ML models. Indeed, our 2\textsuperscript{nd}-order CCNN can pick up this correlation, and only manages to achieve $\approx63\%$ classification accuracy. Further measurements find that all further-range two-point spin-spin correlators are nearly indistinguishable between these two theories.

\begin{figure}[h!]
	\centering
	\includegraphics[width=0.8\linewidth]{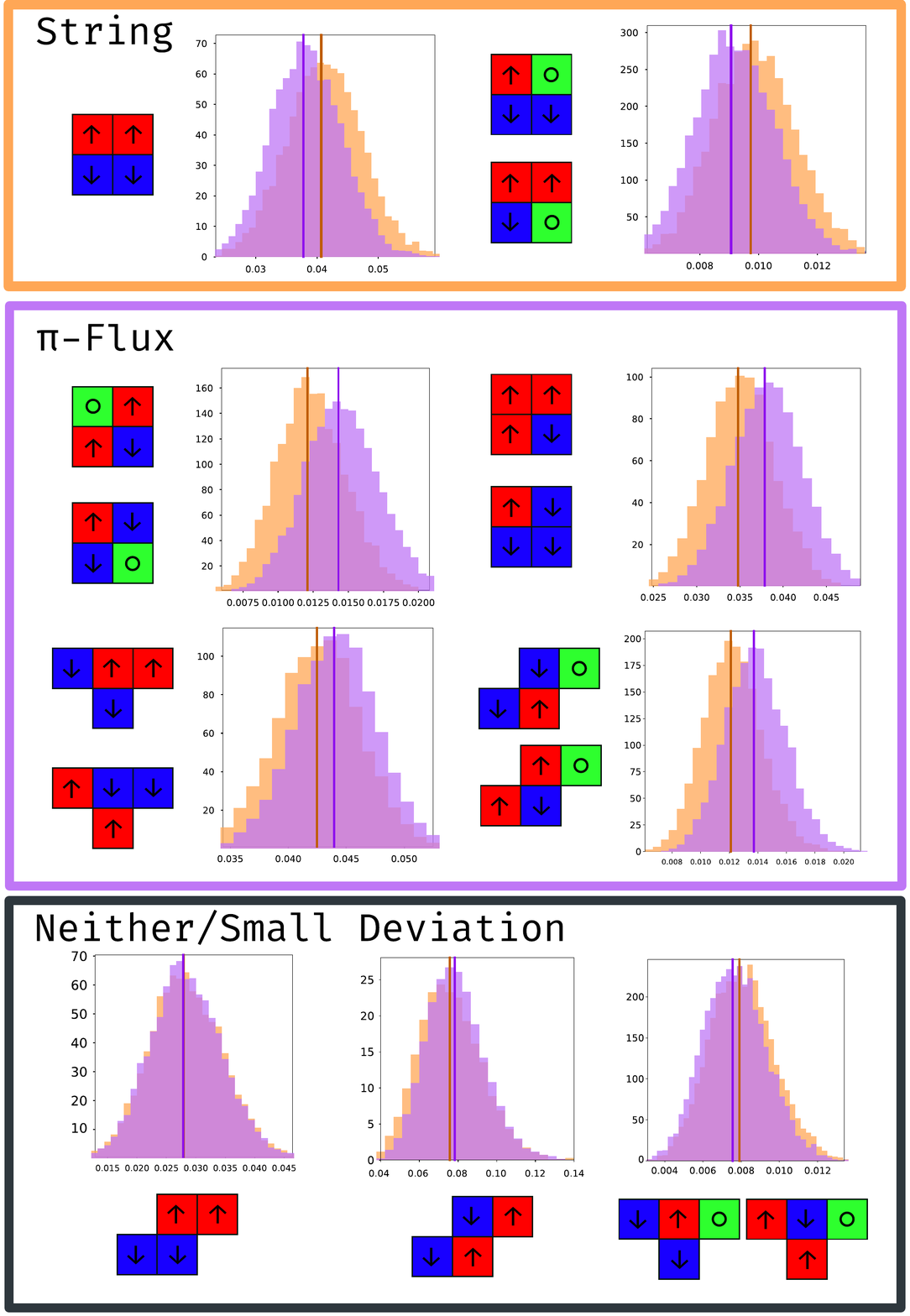}
	\caption{Explicit statistical measurements of the fourth-order correlators discovered in Fig.\ \ref{fig_expansion}. Histogrammed are normalized counts of each pattern (and its symmetry equivalents) obtained from single snapshots of each theory, with each histogram scaled to integrate to one. Vertical lines denote the mean of each distribution.}
	\label{fig_supp_measurements}
\end{figure}

We now turn to the fourth-order correlations discovered by the CCNN. In Fig.~\ref{fig_supp_measurements}, we show histograms of correlator estimates obtained from single snapshots contained within the two datasets. Due to the $D^8$ symmetry of the models, we average over all symmetry-equivalent versions of each correlator for each estimate as the symmetrization of our ML models would. From the figure, we can see that the patterns which are the dominant subpatterns of the learned CCNN filters are indeed biased towards the theory in alignment with what the model predicts, with many distributions being more clearly separated than the two-point NN correlator distributions from Fig.~\ref{fig_supp_simple_nncorr}. We can also see from the figure that some subpatterns contained in the filters actually show no significant difference between the two theories; our interpretation of this is that these patterns emerge as ``connections'' when the CCNN attempts to include multiple significant patterns within a single filter. Since these connecting patters are statistically identical between the two theories, including them is a ``free'' action to the network which will not hamper performance.

\begin{figure}[h!]
	\centering
	\includegraphics[width=0.8\linewidth]{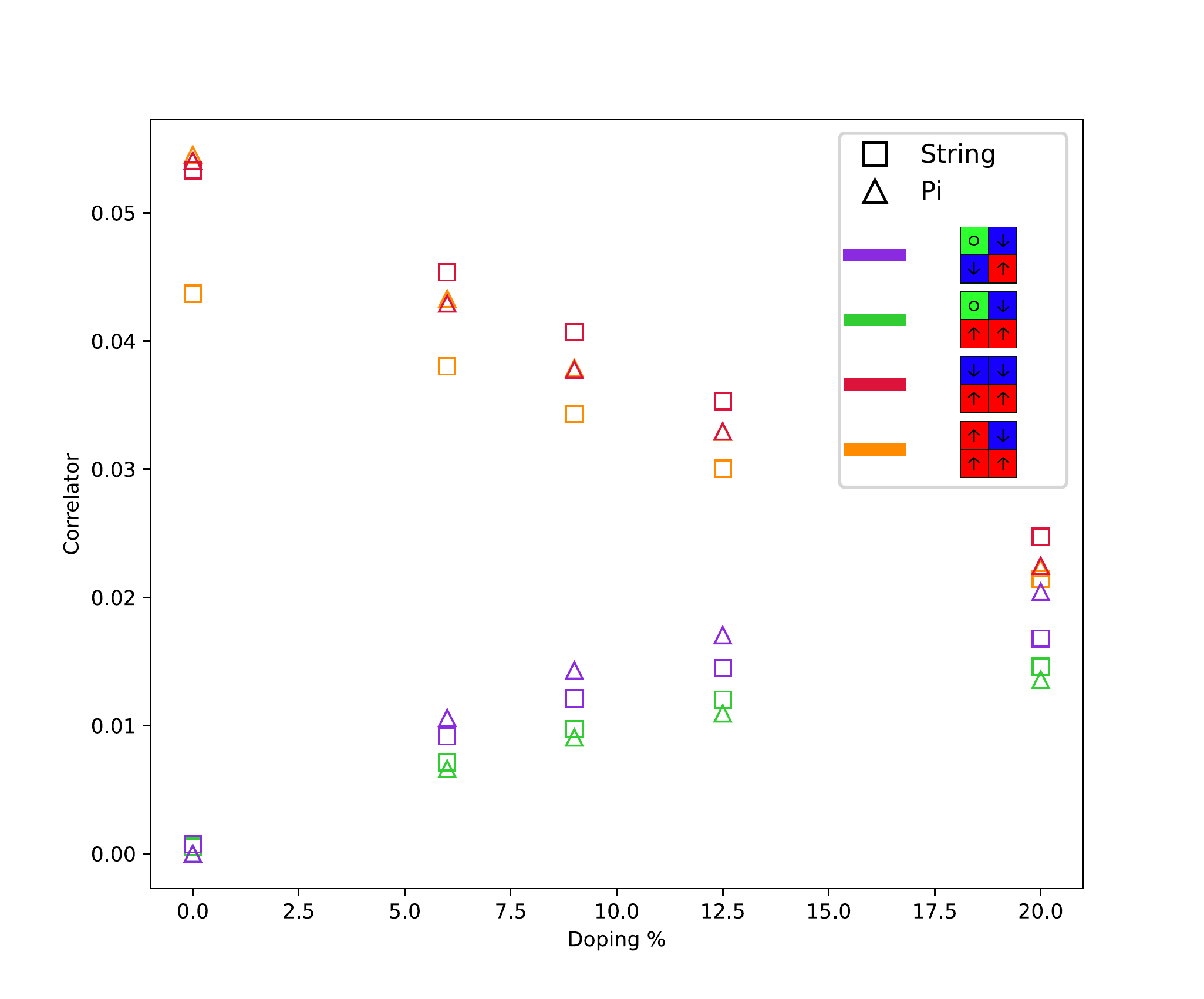}
	\caption{Measured correlators from the snapshot datasets of both theories as a function of hole doping. This work trained on data sampled at 9\% doping.}
	\label{fig_supp_measurements_doping}
\end{figure}

In Fig.\ \ref{fig_supp_measurements_doping}, we plot measured fourth-order correlators obtained from the two datasets as a function of hole doping. While all models in this work are trained on data at 9\% doping, this plot shows an interesting trend. We note that at 0\% doping, the ``parallel spin'' correlator (red) is nearly identical between the two theories. It is only once a finite hole doping is introduced that these correlators begin to deviate from each other. This agrees with our explanation of strings leaving a ``wake'' of parallel spins, increasing this four-site correlator relative to the $\pi$-flux theory.

\newpage
\subsection{S.V Regularization Paths of Traditional CNNs}

The regularization path technique demonstrated in the main text can also be applied to shallow traditional CNNs as an alternative to more complex techniques such as Layerwise Relevancy Propagation \cite{bach_pixel-wise_2015} which are appropriate for deeper models. In these scenarios, regularization paths can still determine which \textit{filters} are contributing most to the classification, however information about which \textit{orders} of correlations are important is inaccessible. We present here an example, and additionally show that fully-connected layers tend to simply ``memorize'' quantum gas data rather than learn true physics by ``reducing'' a full CNN architecture to a simpler one with minimal modifications

\begin{figure}[h!]
    \begin{minipage}{0.45\columnwidth}
        \subfigure[A diagram of the full CNN adapted from \cite{bohrdt_classifying_2019}.]{
            \includegraphics[width=\linewidth]{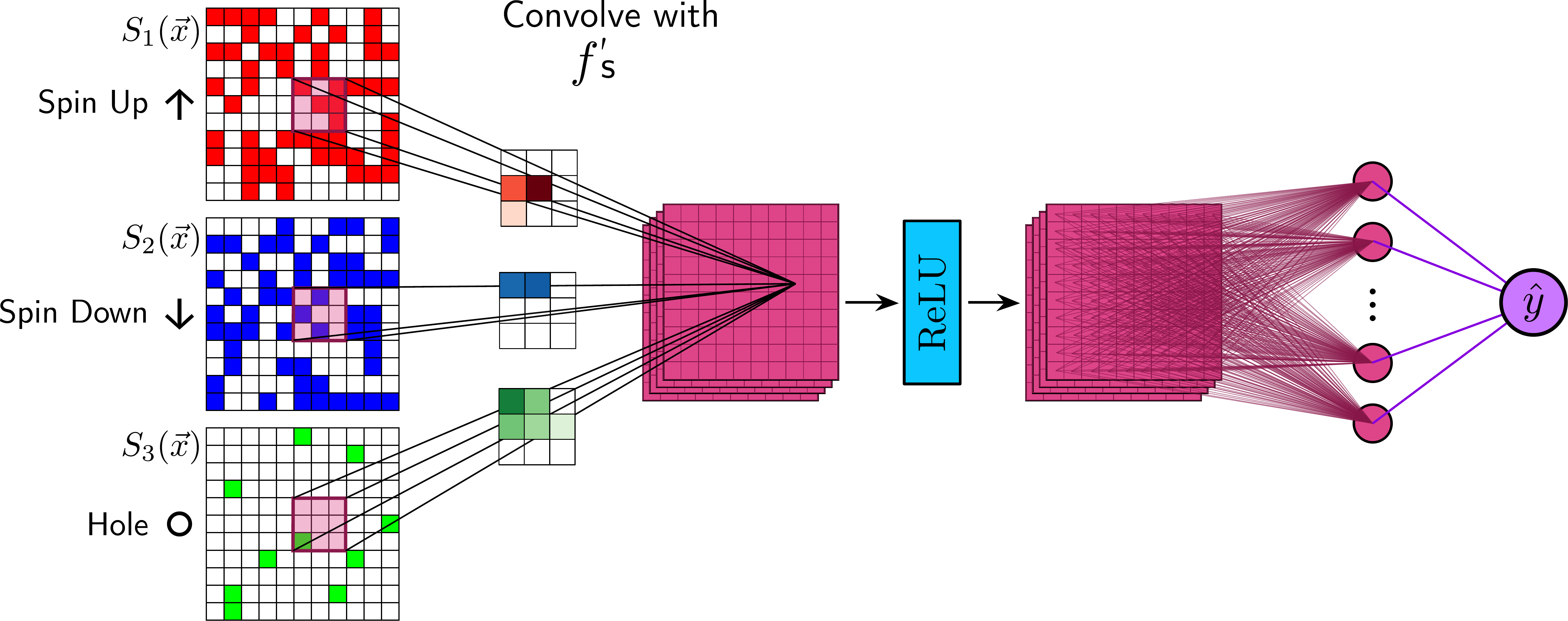}
        }
        \subfigure[A diagram of the reduced CNN.\label{fig_supp_reduced_arch}]{
            \includegraphics[width=\linewidth]{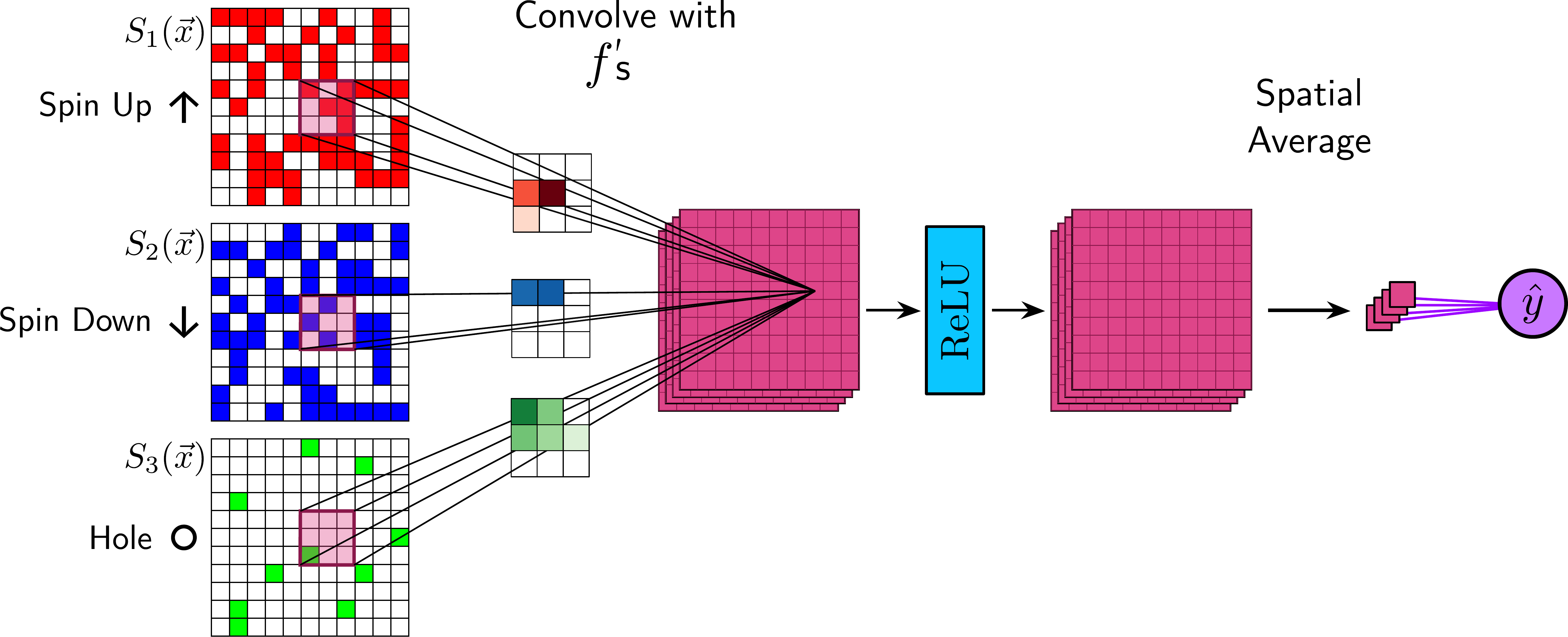}
        }
    \end{minipage}\hfill
    \begin{minipage}{0.45\columnwidth}
        \subfigure[Regularization path obtained from reduced CNN]{
            \includegraphics[width=\linewidth]{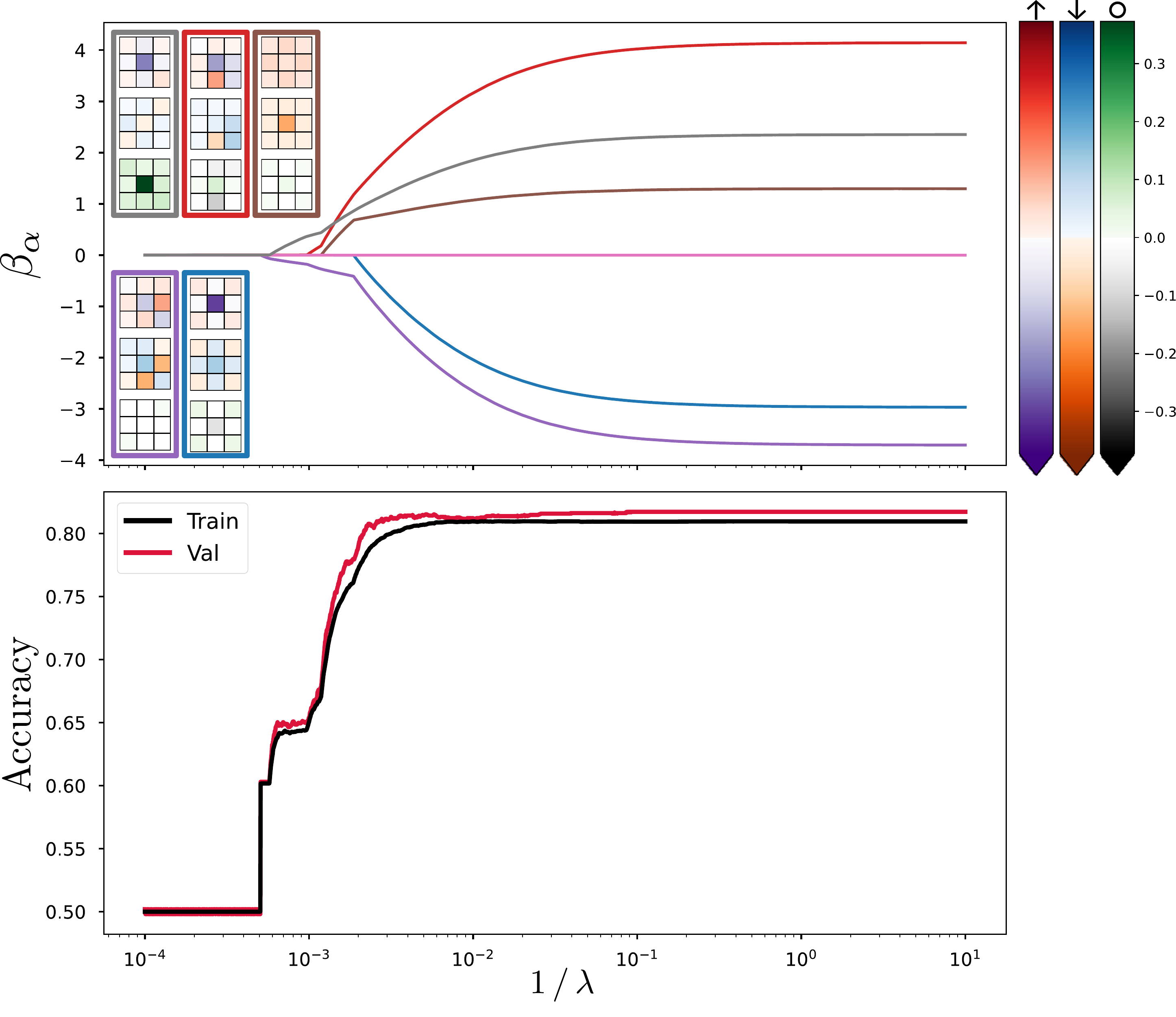}
        }
    \end{minipage}
    \caption{Attempting a regularization path procedure on a standard CNN architecture}
    \label{fig_supp_cnn_path}
\end{figure}

As a demonstration, consider a CNN with an extremely similar architecture as our CCNN, but using a standard nonlinear step, as seen in Fig.~\ref{fig_supp_reduced_arch}. The input $S$ is convolved with a set of learned filters $f_\alpha$ to produce activation maps $C_\alpha$. Then, a nonlinear function $\sigma$ is applied pixelwise to the activation map to produce $\tilde{C}_\alpha = \sigma(C_\alpha)$; here we choose $\sigma = \relu$. These are spatially averaged to produce $c_\alpha$ features, which are used by a final logistic classifier with coefficients $\beta_\alpha$. Heuristically, each $c_\alpha$ captures how much the patterns seen in the snapshot $S$ ``look like'' the filter patterns $f_\alpha$. Note that a BatchNorm layer is not strictly needed here due to the relatively uniform scale of the features $c_\alpha$; it may help training progress easier, but is not required for the model to work. Hence, we will omit it for this test. Additionally, we have found that the model becomes extremely difficult to train if the filter parameters $f_\alpha$ are forced positive as we did for CCNNs. As a consequence, we were unable to apply this constraint, resulting in filters which are harder to interpret than for CCNNs.

To perform a ``reduction'' to this model from the architecture of \cite{bohrdt_classifying_2019}, we first train their architecture exactly as we did for CCNNs: the full model is first trained with an L1 regularization on the filters $f_\alpha$. After training, the filters are frozen, the fully connected layer is replaced with simple spatial averaging, and the $\beta_\alpha$ coefficients are regressed at various strengths of regularization applied to them. The result of doing this process can be seen in \ref{fig_supp_cnn_path}. While not shown, we observed that the replacement of the fully-connected layer with simple averaging actually \textit{improves} validation performance: this provides empirical evidence that this overparameterized layer is simply ``memorizing'' the input data rather than learning true physical features.

We can attempt to interpret the path in Fig.\ \ref{fig_supp_cnn_path} by examining the patterns of the filters which activate first. The first feature to activate, labeled in purple, seems to have something to do with local antiferromagnetic correlations. The second feature, in grey, as well as the blue feature, seem to have something to do with spin-hole correlations. However, it is not as clear how we should understand the red or brown features which activate. Additionally, for each of these patterns we are unable to tell what order feature from the pattern is actually used. The additional complexity of allowing for negative values in filters, along with being unable to disentangle different orders of correlations, make direct interpretation of traditional CNNs generally difficult.

\subsection{S.VI ``Debugging'' Trivial Behaviors}

As a demonstration of a case where CCNNs can be helpful in detecting ``trivial'' features of the data, we will train a CCNN to distinguish between real experimental data and simulated $\pi$-flux snapshots, both at $9\%$ doping. We tune the temperature of the $\pi$-flux simulation as to best match the nearest-neighbor spin correlator with experimental data, as done in \cite{bohrdt_classifying_2019}. The snapshots are zero-masked to the same geometry of the experiment. Because our QGM experiments currently cannot resolve both spin species simultaneously, we convert all spin-down sites in the $\pi$-flux snapshots to appear as empty sites. Additionally, double-hole pairs also appear as empty sites in the experiment due to a parity projection; to accomodate this we insert doublon-hole pairs randomly on neighboring sites in the lattice with a probability matching the theoretical predictions, also appearing as empty sites. As we have two species of site (spin up and ``empty''), this results in a set of two-channel snapshots input to the network.

We only have access to a small amount of experimental data, meanwhile $\pi$-flux data is essentially limitless. However, if we naively train a ML model with a dataset that contains significantly more data of one class than the other, a significant local minima the network can be trapped in is to simply predict the more populous class all the time. For example, in our dataset we have $2476$ experimental snapshots, and $19500$ $\pi$-flux snapshots. A model which predicts $\pi$-flux all the time would achieve an $89\%$ accuracy! A simple solution to this is to \textit{oversample} the experimental snapshots. On each epoch, we randomly duplicate snapshots from the experimental dataset until there are $19500$ in the dataset.

\begin{figure}[h!]
    \subfigure[Filters learned from a model trained to classify experimental v. $\pi$-flux data \label{fig_supp_exppi_filts}]{
        \includegraphics[width=0.4\columnwidth]{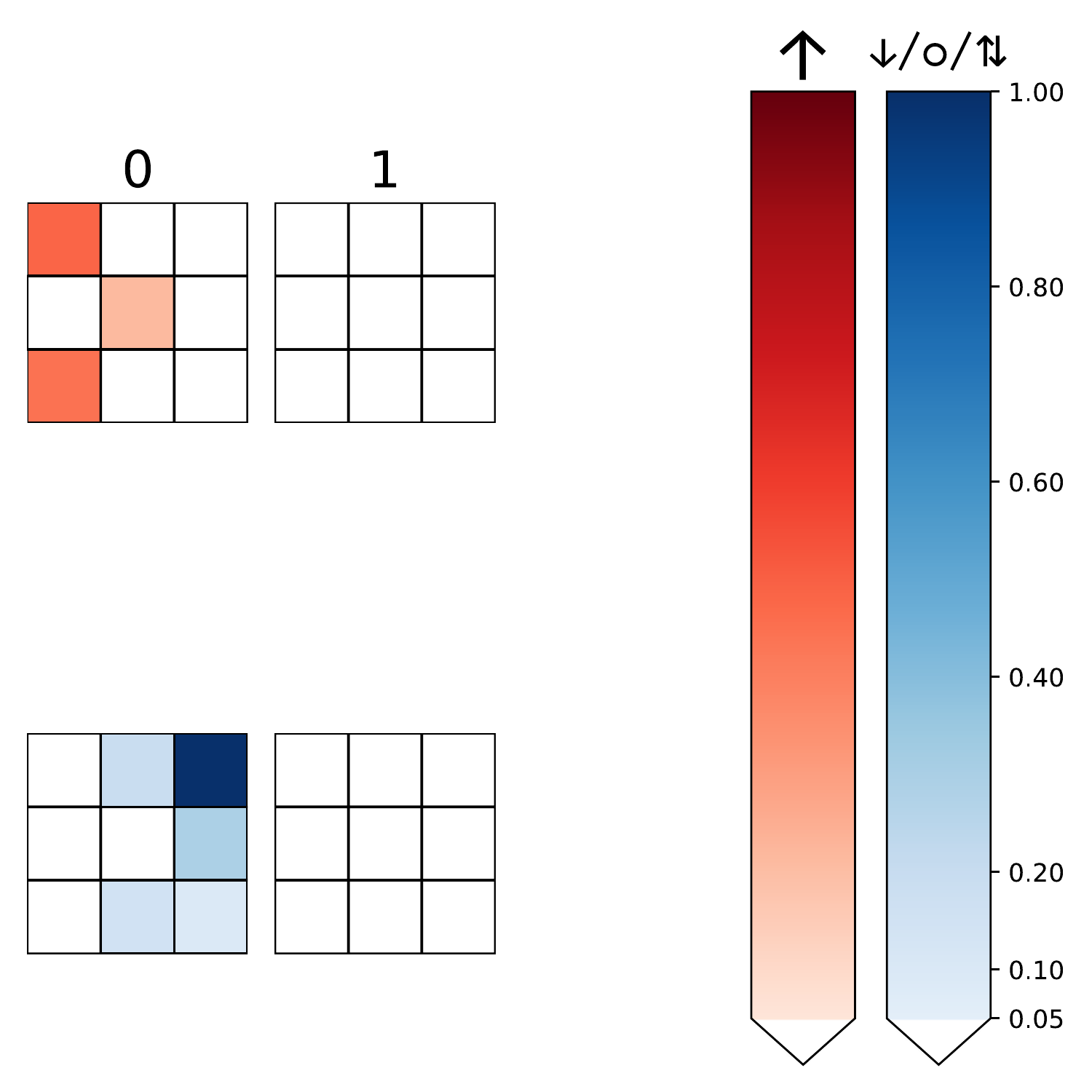}
    } \hfill
    \subfigure[Regularization path related to filters shown in \ref{fig_supp_exppi_filts}. Positive coefficients correspond to $\pi$-flux snapshots, negative correspond to experiment. \label{fig_supp_exppi_path}]{
        \includegraphics[width=0.4\columnwidth]{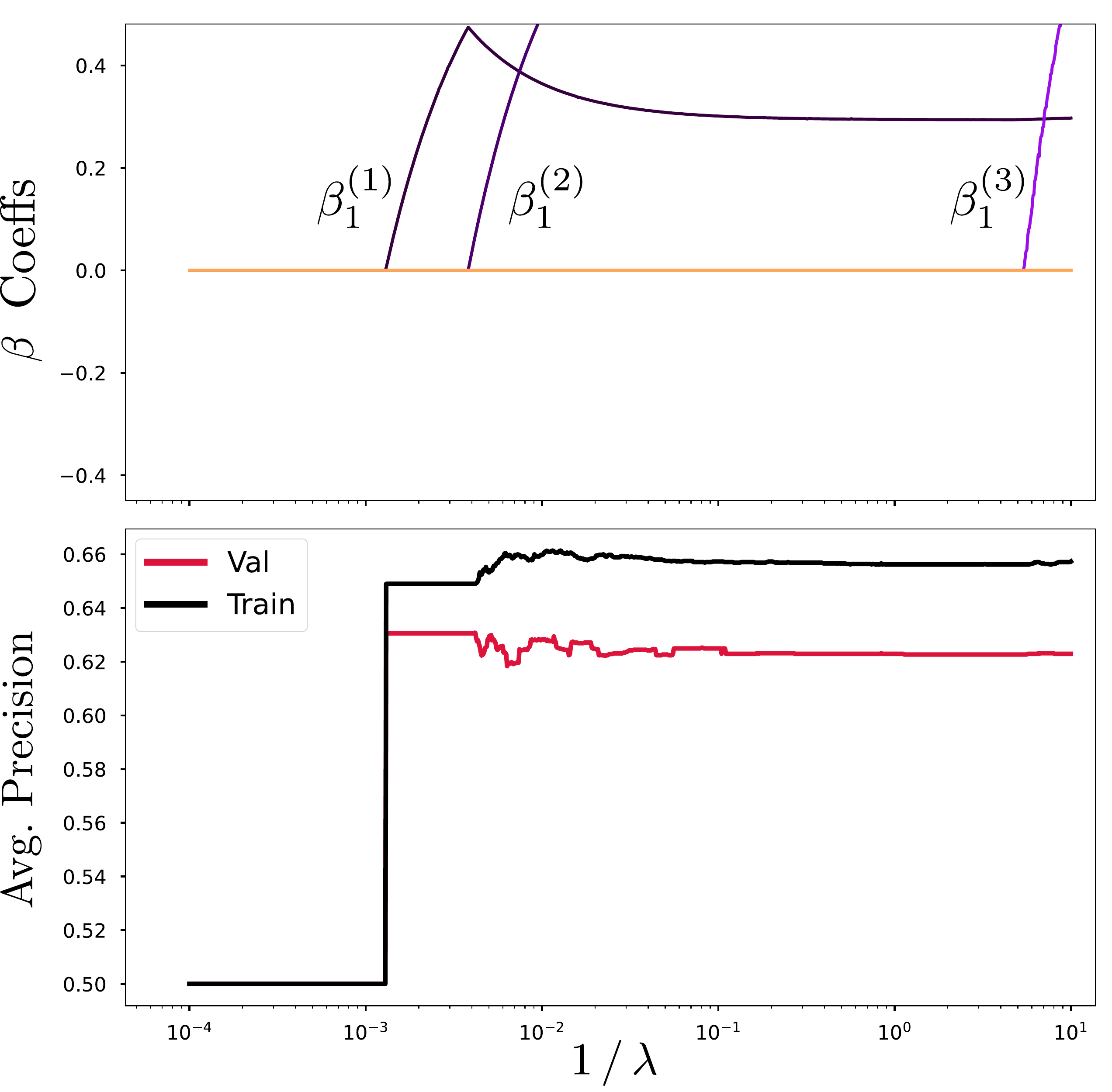}
    }
    \caption{Model behavior when classifying experimental v. $\pi$-flux data. Here, due to the class imbalance, we plot average precision rather than accuracy in the regularization path, which can be thought of as the average of the two accuracies on the classes individually.}
\end{figure}

As in the main text, we train a CCNN with two filters, and with an L1 loss applied to the filter weights. We use a slightly stronger value of $\lambda = 0.01$, and this was found to not significantly reduce final accuracy compared to the no-regularization case. (Note that the accuracy is much lower here mainly due to the smaller spatial extent of the snapshot: the circular region accessed by experiment contains only $90$ sites while the snapshots in the main text contain $15\times15 = 225$). The final filters learned are shown in Fig.\ \ref{fig_supp_exppi_filts}. We can see that the L1 loss has completely turned off one of the filters, while the other still has a collection of pixels left on.

We might originally guess that this is looking for specific patterns in the snapshots that somewhat resemble the pattern of the filter, i.e. correlators which are subpatterns. However, once constructing the regularization path from the model, shown in Fig.~\ref{fig_supp_exppi_path}, we can see this is not the case! The $1$st order correlator explains essentially all of the network's performance, meaning the network is really just measuring the ``empty site'' occupancy. (In fact, we can see in this instance the higher-order correlations contribute to overfitting, as the validation precision drops when they activate). A simple explanation of this is that the doublon-hole density in the experiment must be higher than expected for the given temperature.

\newpage

\subsection{S.VII Further Explanation of the CCNN operation}

\begin{figure}[h!]
	\centering
	\includegraphics[width=\linewidth]{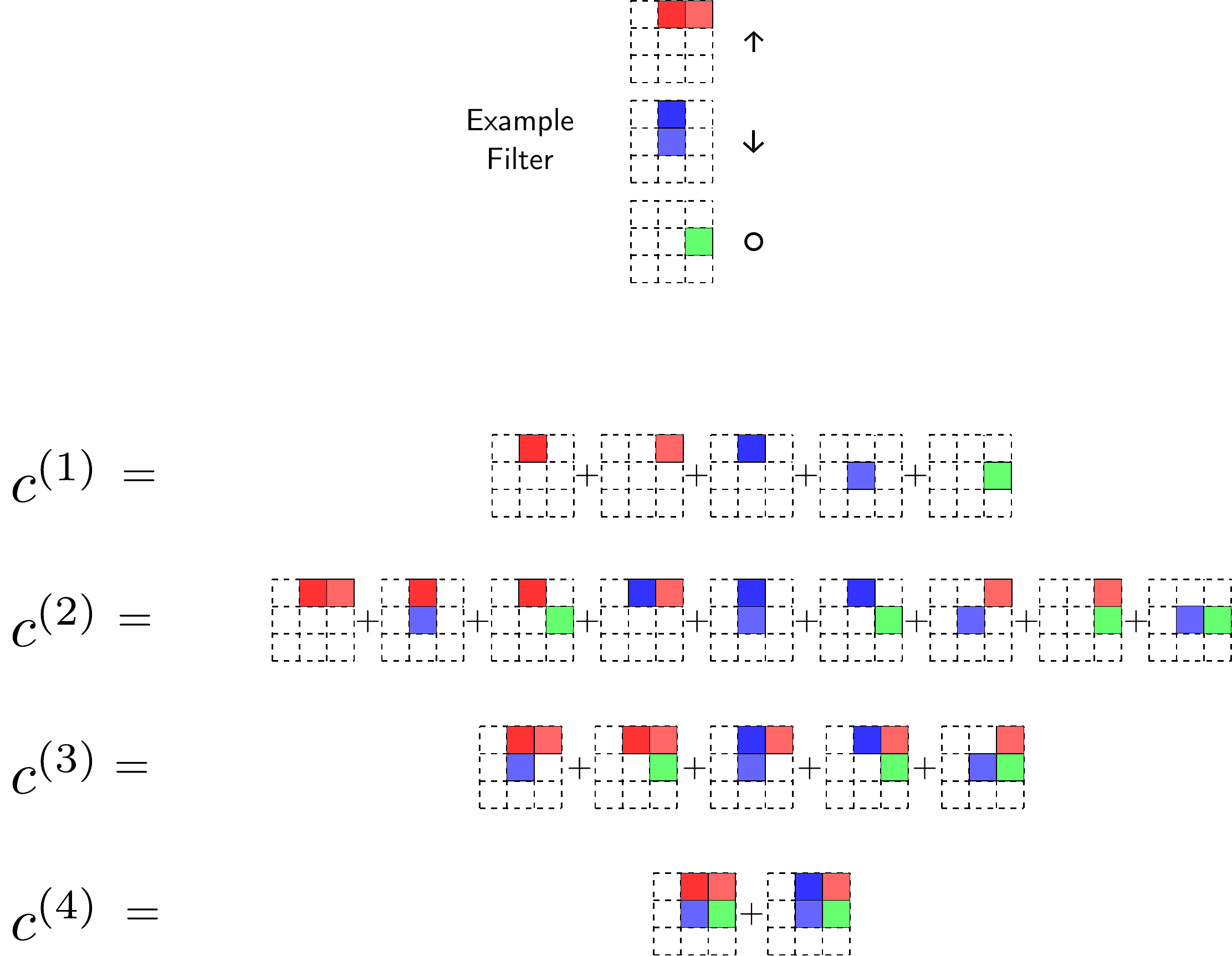}
	\caption{An explanation of the nonlinear features $c^{(n)}$ in terms of multi-site correlators for an example filter. White pixels in the filter are zero weight. Each term in the expansion can be understood as counting the number of occurrences of the shown pattern in the snapshot, weighted by the product of the intensities of the pixels comprising the pattern.}
	\label{fig_drawing_corr}
\end{figure}

In Fig.~\ref{fig_drawing_corr}, we explicitly write out the terms of Eq.~\ref{eq_Fn} for an example filter. We can see that for the feature $C^{(n)}$, to find all the terms one draws all patterns which can be made by choosing $n$ pixels from across the channels. As the Hilbert space of each of our models is restricted to the singly-occupied subspace, we do not need to consider patterns with more than one pixel at the same site. Each of these terms is weighted by the product of the intensities of the pixels which constitute the pattern.

\end{document}